\documentclass[a4paper,11pt]{article}
\pdfoutput=1 

\usepackage{jcappub}
\usepackage[T1]{fontenc}
\usepackage[utf8]{inputenc}
\usepackage{physics} 

\usepackage{natbib}

\newcommand{\bea}{\begin{eqnarray}} \newcommand{\eea}{\end{eqnarray}}
\newcommand{\reh}{{\mr{reh}}}

\newcommand{\re}[1]{(\ref{#1})}
\newcommand{\PR}{\mathcal{P}_\mathcal{R}}
\newcommand{\PRfrozen}{\mathcal{P}_{\mathcal{R}\infty}}
\newcommand{\PT}{\mathcal{P}_\mathcal{T}}
\newcommand{\R}{\mathcal{R}}
\newcommand{\mr}{\mathrm}
\newcommand{\Mpl}{M_\mr{Pl}}

\title{Planck scale black hole dark matter from Higgs inflation}

\author[a,b]{Syksy~R{\"a}s{\"a}nen}
\author[a]{and Eemeli~Tomberg}

\affiliation[a]{University of Helsinki, Department of Physics and Helsinki Institute of Physics,\\ P.O. Box 64, FIN-00014 University of Helsinki, Finland}

\affiliation[b]{Birzeit University, Department of Physics \\
P.O. Box 14, Birzeit, West Bank, Palestine}

\emailAdd{syksy.rasanen@iki.fi}
\emailAdd{eemeli.tomberg@helsinki.fi}

\abstract{We study the production of primordial black hole (PBH) dark matter in the case when the Standard Model Higgs coupled non-minimally to gravity is the inflaton. PBHs can be produced if the Higgs potential has a near-critical point due to quantum corrections. In this case the slow-roll approximation may be broken, so we calculate the power spectrum numerically. We consider both the metric and the Palatini formulation of general relativity.
Combining observational constraints on PBHs and on the CMB spectrum we find that PBHs can constitute all of the dark matter only if they evaporate early and leave behind Planck mass relics. This requires the potential to have a shallow local minimum, not just a critical point. The initial PBH mass is then below $10^6$ g, and predictions for the CMB observables are the same as in tree-level Higgs inflation, $n_s=0.96$ and $r=5\times10^{-3}$ (metric) or $r=4\times 10^{-8} \dots 2 \times 10^{-7}$ (Palatini).}

\begin{document}

\begin{flushleft}
	\hfill		  HIP-2018-23/TH \\
\end{flushleft}

\maketitle
\flushbottom

\section{Introduction} \label{sec:intro}

\paragraph{Higgs inflation.}

The Standard Model (SM) of particle physics is very successful in describing physics at the electroweak (EW) scale. A key part of the SM is the Higgs mechanism, in which the Higgs field gives masses to elementary particles. If the Higgs field is coupled non-minimally to gravity and the theory is extrapolated to high energies, the tree-level theory provides a model of inflation \cite{Bezrukov:2007ep, Bezrukov:2013fka}. Its predictions for the spectral index of primordial scalar perturbations and tensor-to-scalar ratio are in excellent agreement with observations of the cosmic microwave background (CMB) radiation and large-scale structure \cite{Akrami:2018odb}. Higgs inflation is appealing in its simplicity, and it is perhaps the most minimalistic way of incorporating inflation into known particle physics.

However, the predictions can change because of two complicating factors: choice of gravitational degrees of freedom and quantum corrections. There are several formulations of general relativity that are equivalent for the Einstein--Hilbert action and minimally coupled matter. The most common is the metric formulation, where the metric and its derivatives are the only degrees of freedom. Another one is the Palatini formulation, where the connection is taken to be an independent variable \cite{Einstein:1925, Ferraris:1982}, which leads to different predictions when the gravitational action is more complicated than the Einstein--Hilbert case \cite{Buchdahl:1960, Buchdahl:1970, ShahidSaless:1987, Flanagan:2003a, Flanagan:2003b, Sotiriou:2006, Sotiriou:2008, Olmo:2011, Borunda:2008, Querella:1999, Cotsakis:1997, Jarv:2018, Conroy:2017, Li:2007, Li:2008, Exirifard:2007, Afonso:2017, Afonso:2018c, Enckell:2018hmo, Antoniadis:2018ywb}, or when the matter is directly coupled to the Ricci scalar \cite{Lindstrom:1976a, Lindstrom:1976b, Bergh:1981, Bauer:2008, Bauer:2010jg, Koivisto:2005, Rasanen:2017ivk, Enckell:2018kkc, Markkanen:2017, Kozak:2018, Jarv:2017}, as in Higgs inflation \cite{Bauer:2008, Bauer:2010jg, Rasanen:2017ivk, Enckell:2018kkc, Markkanen:2017}.

Quantum corrections can in principle provide a consistency test between cosmology and collider physics \cite{Espinosa:2007qp, Barvinsky:2008ia, Barvinsky:2009fy, Burgess:2009ea, Popa:2010xc, DeSimone:2008ei, Bezrukov:2008ej, Bezrukov:2009db, Barvinsky:2009ii, Bezrukov:2010jz, Bezrukov:2012sa, Allison:2013uaa, Salvio:2013rja, Shaposhnikov:2013ira}, but they bring up ambiguities related to the non-renormalizable nature of gravity and possibly to the choice of frame \cite{Weenink:2010rr, Calmet:2012eq, Steinwachs:2013tr, Prokopec:2014iya, Kamenshchik:2014waa, Burns:2016ric, Fumagalli:2016lls, Hamada:2016, Karamitsos:2017elm, Karamitsos:2018lur, Bezrukov:2008ej, Bezrukov:2009db, Bezrukov:2010jz, Allison:2013uaa, George:2013iia, Postma:2014vaa, Prokopec:2012ug, Herrero-Valea:2016jzz, Pandey:2016jmv, Pandey:2016unk}, lead to possible issues with unitarity \cite{Barbon:2009ya, Burgess:2009ea, Hertzberg:2010dc, Bauer:2010jg, Bezrukov:2010jz, Bezrukov:2011sz, Calmet:2013hia, Weenink:2010rr, Prokopec:2012ug, Xianyu:2013rya, Prokopec:2014iya, Ren:2014sya, Escriva:2016cwl}, and may make the Higgs self-coupling run to a negative value at high energies \cite{Espinosa:2015qea, Espinosa:2015kwx, Iacobellis:2016eof, Butenschoen:2016lpz, Espinosa:2016nld}. Nevertheless, thanks to a conjectured approximate shift symmetry at the inflationary scale, loop corrections there can be calculated systematically independent of any problems at lower scales, though the connection between the inflationary regime and EW scale physics is not uniquely defined \cite{Bezrukov:2010jz, George:2013iia, Calmet:2013hia, Bezrukov:2014bra, Bezrukov:2014ipa, Rubio:2015zia, George:2015nza, Saltas:2015vsc, Fumagalli:2016lls, Enckell:2016xse, Moss:2014nya, Bezrukov:2017dyv, Enckell:2018kkc}. In most cases, quantum corrections don't affect the CMB predictions much \cite{Fumagalli:2016lls}. However, if the couplings are tuned, the corrections can produce a feature at the inflationary scales, such as a critical point where the first and second derivatives of the potential vanish (simply called an inflection point in some publications) \cite{Allison:2013uaa, Bezrukov:2014bra, Hamada:2014iga, Bezrukov:2014ipa, Rubio:2015zia, Fumagalli:2016lls, Enckell:2016xse, Bezrukov:2017dyv, Rasanen:2017ivk, Salvio:2017oyf, Masina:2018}, a hilltop \cite{Fumagalli:2016lls, Rasanen:2017ivk, Enckell:2018kkc} or a degenerate vacuum \cite{Jinno:2017jxc, Jinno:2017lun}. Such features can change the dynamics and the CMB predictions considerably.

\paragraph{Primordial black holes as dark matter.}

In addition to affecting CMB observables, features in the potential could produce large scalar perturbations that seed primordial black holes (PBHs) \cite{Novikov:1979} when scales around the critical point re-enter the Hubble radius after inflation. PBHs could then constitute part or all of the dark matter \cite{Chapline:1975, Dolgov:1992pu, Jedamzik:1996mr, Ivanov:1994pa, GarciaBellido:1996qt, Yokoyama:1995ex, Ivanov:1997ia, Blais:2002nd}. The recent detection of gravitational waves from black hole mergers by LIGO and Virgo \cite{TheLIGOScientific:2016pea} has rekindled the study of PBHs \cite{Carr:1974nx, Carr:1975qj}, with many proposed models of critical point inflation \cite{Garcia-Bellido:2017mdw, Ezquiaga:2017fvi, Kannike:2017bxn, Germani:2017bcs, Motohashi:2017kbs, Gong:2017qlj, Ballesteros:2017fsr} that could produce PBH dark matter in the mass range observable by current or upcoming gravitational wave experiments. Such predictions are subject to stringent constraints from astrophysical observations, such as lensing and gamma-ray bursts \cite{Carr:2009jm, Carr:2017jsz, Carr:2016drx}. Only four regions remain in the spectrum of PBH mass that still allow a sizeable PBH population. There are two narrow mass windows at $10^{18}$ g and $4\times10^{19}$ g, a mass window at $10^{34}\dots10^{35} \, \mr{g} \approx 25\dots100 \, \mr{M}_\odot$ close to the LIGO/Virgo range, and finally all initial masses $<10^6$ g are allowed, corresponding to PBHs that evaporate before big bang nucleosynthesis (BBN) down to Planck scale relics without spoiling baryogenesis \cite{Carr:2009jm, Carr:2017jsz}.

PBH production in critical point Higgs inflation has been studied in a simple setting in \cite{Ezquiaga:2017fvi}. The authors concluded that with a suitable choice of parameters, their model can produce enough PBHs in the LIGO/Virgo mass range to constitute all of the dark matter while remaining consistent with CMB observations. However, their analysis was based on the slow-roll (SR) approximation, which does not necessarily apply at a critical point \cite{Downes:2012, Germani:2017bcs, Motohashi:2017kbs, Gong:2017qlj, Kannike:2017bxn, Ballesteros:2017fsr}. Also, the running of the quartic Higgs coupling and of the non-minimal coupling was treated phenomenologically without using the full renormalisation group equations.

In this study of critical point Higgs inflation, we start from the renormalization group improved SM Higgs potential and run it from the EW scale up. We also run the chiral SM down from high scales and match the two potentials at an intermediate scale, with a jump in the quartic Higgs coupling and the top Yukawa coupling. We scan over all near-critical point Higgs potentials and calculate the spectrum of scalar perturbations numerically, without the SR approximation. We consider both the metric and the Palatini formulation of general relativity. We compare to the observational limits on the CMB spectrum and on PBHs.

In section \ref{sec:higgs_pot} we outline how we find the potential with a critical point. In section \ref{sec:inflation} we discuss inflation and SR violation near such a critical point. In section \ref{sec:PBHs} we summarize the theory of primordial black holes and their formation in the context of critical point inflation. In section \ref{sec:scans} we present our numerical scans over all allowed critical and near-critical point potentials and the results for black hole formation in Higgs inflation. Section \ref{sec:discussion} is reserved for discussion, and in section \ref{sec:conclusions} we summarize our findings. Technical details about the mass limit of relic PBHs and non-conservation of the curvature perturbation are presented in appendices \ref{app:uppermasslimit} and \ref{app:PRgrowth}.

\section{Higgs potential} \label{sec:higgs_pot}

\subsection{Tree-level potential} \label{sec:tree_level_pot}

The Lagrangian of the SM coupled non-minimally to gravity is
\begin{equation} \label{eq:higgs_SM_action}
	S = \int d^4 x \sqrt{-g} \qty[ \frac{1}{2}\left(M^2 + \xi h^2\right) g^{\mu\nu} R_{\mu\nu} - \frac{1}{2} g^{\mu\nu} \partial_\mu h \partial_\nu h - V(h) + \mathcal{L}_{SM} ] \ ,
\end{equation}
where $g_{\mu\nu}$ is the metric, $R_{\mu\nu}$ is the Ricci tensor, $M$ is a mass scale close to Planck mass that we set to unity henceforth, $h$ is the radial Higgs field, $\xi$ is the non-minimal coupling, $V(h) = \frac{\lambda}{4}h^4$, and $\mathcal{L}_{SM}$ contains the rest of the SM. The Ricci tensor is built from the connection: in the metric case we take it to be the Levi--Civita connection, but in the Palatini case it is an independent variable determined by the field equations \cite{Bauer:2008}.
To make contact with the usual analysis of inflation, it is customary to perform a Weyl transformation to the Einstein frame and define a new canonical scalar field $\chi$ with minimal coupling to gravity and a canonical kinetic term \cite{Bezrukov:2007ep}:
\begin{equation} \label{eq:weyl_transform}
	g_{\alpha\beta}\rightarrow (1+\xi h^2)^{-1} g_{\alpha\beta} \ , \qquad \frac{d h}{d \chi} = \frac{1 + \xi h^2}{\sqrt{1 + \xi h^2  + p 6\xi^2 h^2}} \ ,
\end{equation}
where $p=1$ in the metric formulation and $p=0$ in the Palatini formulation. In the metric case the Einstein frame potential is
\begin{equation} \label{eq:higgs_pot_einstein}
	U(\chi) = \frac{V[h(\chi)]}{[1+\xi h(\chi)^2]^2} \equiv
	 \frac{\lambda}{4}F[h(\chi)]^4 \ ,
\end{equation}
where in the metric case
\begin{equation} \label{eq:F_metric}
	F(h) \equiv \frac{h}{\sqrt{1+\xi h^2}} \approx \begin{cases}
    \chi & \qquad h \ll 1/\xi \\
    \frac{1}{\sqrt{\xi}}\qty(1-e^{-\sqrt{2/3}\chi})^{1/2} & \qquad h \gg 1/\sqrt{\xi} \ .
  \end{cases}
\end{equation}
In the Palatini case, we have the exact result
\begin{equation} \label{eq:F_palatini}
	F[h(\chi)] = \frac{1}{\sqrt{\xi}}\tanh \qty(\sqrt{\xi}\chi) \ .
\end{equation}
It is convenient to define \cite{Fumagalli:2016lls}
\begin{equation} \label{eq:delta}
	\delta \equiv \frac{1}{\xi h^2} \ .
\end{equation}
The potential \eqref{eq:higgs_pot_einstein} is exponentially flat and supports SR inflation for $h \gtrsim 1/\sqrt{\xi}$, that is, $\delta \lesssim 1$.

\subsection{Quantum corrections} \label{sec:quantum_corrections}

We take into account quantum corrections to the potential in the same way as in \cite{Enckell:2018kkc}. At $\delta \lesssim 1$ the theory is approximated by the chiral SM, and we add to \eqref{eq:higgs_pot_einstein} the chiral SM one-loop potential correction \cite{Bezrukov:2009db}
\begin{equation} \label{eq:one_loop_corr}
	U_{\mathrm{1-loop}} = \frac{6m_W^4}{64\pi^2}\left(\ln \frac{m_W^2}{\mu^2} - \frac{5}{6} \right) +
				\frac{3m_Z^4}{64\pi^2}\left(\ln \frac{m_Z^2}{\mu^2} - \frac{5}{6} \right) -
				\frac{3m_t^4}{16\pi^2}\left(\ln \frac{m_t^2}{\mu^2} - \frac{3}{2} \right) \ ,
\end{equation}
where $W$ and $Z$ boson masses $m_W$ and $m_Z$ and the top quark mass $m_t$ are
\begin{equation} \label{eq:large_masses}
	m_W^2 = \frac{g^2 F^2}{4} \ , \qquad m_Z^2 = \frac{\qty(g^2+g'^2) F^2}{4} \ , \qquad m_t^2 = \frac{y_t^2 F^2}{2} \ ,
\end{equation}
and other fermions are approximated to be massless. We also let the couplings run according to the one-loop chiral SM beta functions \cite{Bezrukov:2009db, Dutta:2007st}:
\begin{equation} \label{eq:large_betas}
\begin{alignedat}{2}
	16\pi^2\beta_\lambda &= -6y_t^4 + \frac{3}{8}\qty[2g^4 + (g'^2+g^2)^2] \ , \quad &16\pi^2\beta_{y_t} &= y_t\qty(-\frac{17}{12}g'^2 - \frac{3}{2}g^2 - 8g_S^2 + 3y_t^2) \ , \\
	16\pi^2\beta_{g} &= -\frac{13}{4}g^3 \ , \quad
	16\pi^2\beta_{g'} = \frac{27}{4}g'^3 \ , \quad
	&16\pi^2\beta_{g_S} &= -7g_S^3 \ .
\end{alignedat}
\end{equation} 
A term in $\beta_\lambda$ proportional to $\lambda$ and the running of $\xi$ are omitted as higher order corrections. Everything is done in the $\overline{MS}$ renormalization scheme. In principle, the theory reduces to the chiral SM only in the limit $\delta \rightarrow 0$; however, we assume that the above results give a decent approximation for the potential for all $\delta\lesssim1$. In both the metric formulation and the Palatini formulation the theory is asymptotically the chiral SM, but for finite values of $\delta$ the renormalisation group running is different in the two formulations; our approximation neglects these differences. The renormalization scale is chosen to depend on the field,
\begin{equation} \label{eq:renormscale_chiral}
	\mu(\chi) = \kappa F(\chi) \ ,
\end{equation}
where the constant $\kappa$ is chosen so that the loop correction \eqref{eq:one_loop_corr} vanishes at a feature scale (see below).

We connect the chiral SM to EW scale physics at a threshold scale chosen to be
\begin{equation} \label{eq:jumpscale}
	\mu_1 = \frac{\kappa}{\xi} \ .
\end{equation}
Below this scale, the couplings run according to three loop SM beta functions and we match them to the observed particle masses and strong coupling constant \cite{Aad:2015zhl, Khachatryan:2015hba}
\begin{equation} \label{eq:SM_bestfit_vals}
	\frac{g_S^2(m_Z)}{4\pi}=0.1184 \ , \quad m_H = 125.09 \pm 0.24\,\mathrm{GeV} \ , \quad m_t = 172.44 \pm 0.49\,\mathrm{GeV} \ .
\end{equation}
The top mass uncertainty above does not include the theoretical uncertainty in relating the measured MCMC mass to the perturbation theory pole mass, estimated to be of the order one GeV \cite{Khachatryan:2015hba, Butenschoen:2016lpz}. Running and matching are done with the code \cite{SMRunningCode},  which also uses the initial values $m_W=80.399$ GeV, $m_Z=91.1876$ GeV, the Fermi constant $G_F=1.16637 \times 10^{-5}$ GeV\textsuperscript{$-2$}, the fine structure constant $\alpha=1/127.916$ and $\sin^2 \theta_W=0.23116$, where $\theta_W$ is the Weinberg angle evaluated at scale $m_Z$.
At the threshold, we let couplings $\lambda$ and $y_t$ jump by $\Delta\lambda$ and $\Delta y_t$ to simulate the effects of the unknown physics between the EW and inflationary scale. For simplicity, we omit such jumps for the gauge couplings and match their SM and chiral SM values at the threshold.

\begin{figure}
\begin{center}
\includegraphics{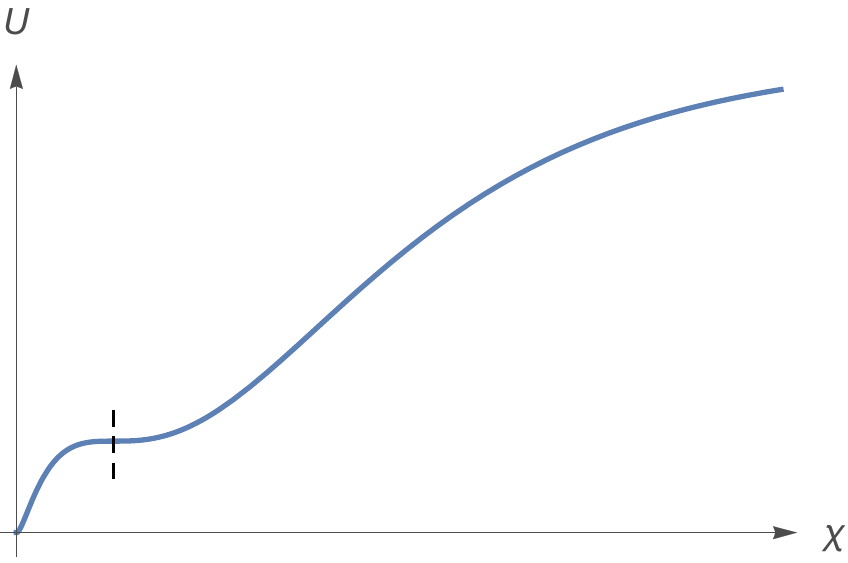}
\end{center}
\caption{Sketch of a critical point potential: at the point marked by the dashed line, $U'=U''=0$. This potential was formed with parameter values $\xi=70$, $\delta_0=1.5$, $\epsilon_{V0}=\eta_{V0}=0$ in the metric formulation (see section \ref{sec:scans}).}
\label{fig:inflection_point_sketch}
\end{figure}

Different values of the jumps  $\Delta\lambda$ and $\Delta y_t$ correspond to different corrections to the potential. We consider a potential with a critical point at the inflationary scales $\delta\lesssim1$, where $U'=U''=0$, see figure \ref{fig:inflection_point_sketch}, or a near-critical point where $U'$ and $U''$ are close to zero.

\section{Inflation} \label{sec:inflation}

\subsection{Background evolution} \label{sec:background}

The potential discussed in section \ref{sec:higgs_pot} gives the time evolution of the homogeneous background field $\chi(t)$ and the scale factor $a(t)$ through the Friedmann equations (taking spatial curvature to be negligible):
\begin{equation}
	\label{eq:friedmann_scalar_eq}
	3 H^2 = \frac{1}{2}\dot{\chi}^2 + U(\chi) \ , \qquad
	\ddot{\chi} + 3H\dot{\chi} + U'(\chi) = 0 \ ,
\end{equation}
where $H\equiv\dot{a}/a$ and dot indicates derivative with respect to the cosmic time $t$. For cosmic inflation to take place, the scalar field potential energy must dominate over the kinetic term, so the first equation in \eqref{eq:friedmann_scalar_eq} simplifies:
\begin{equation} \label{eq:inflation_condition}
	\frac{1}{2}\dot{\chi}^2 \ll U(\chi) \quad \Rightarrow \quad 3 H^2 \approx U(\chi) \ . 
\end{equation}
This condition can be written in terms of the first Hubble SR parameter:
\begin{equation} \label{eq:first_hubble_sr_parameter}
	\epsilon_H \equiv \frac{\dot{\chi}^2}{2 H^2} < 1 \quad \text{for inflation.}
\end{equation}

\subsection{Inflationary observables} \label{sec:observables}

To compare to observations, we have to calculate the evolution of perturbations on top of the background solution. Gaussian perturbations are fully characterized by the power spectrum: $\PR$ for the comoving curvature perturbation in the scalar sector, and $\PT$ for the tensor perturbations. Observations of the CMB temperature and polarization anisotropies and CMB lensing \cite{Akrami:2018odb} give at the pivot scale $k_*=0.05$ Mpc${}^{-1}$ the amplitude
\begin{equation}
	\label{eq:pert_norm}
	A_s \equiv \PR(k_*) = 2.1 \times 10^{-9} \ ,
\end{equation}
spectral index
\begin{equation}
	\label{eq:ns_obs}
	n_s \equiv 1+\frac{d \PR(k)}{d \ln k}\bigg\rvert_{k=k_*} =0.9653\pm0.0041
\end{equation}
and tensor-to-scalar ratio
\begin{equation}
	\label{eq:r_obs}
	r \equiv \frac{\PT(k_*)}{\PR(k_*)} < 0.070 \ .
\end{equation}
The pivot scale $k_*$ exits the Hubble radius when the number of e-folds until the end of inflation is
\begin{equation} \label{eq:Npivot}
\begin{split}
  N_* \approx 61 - \Delta N_\reh + \frac{1}{4}\ln U_*
  \approx 57 + \frac{1}{4}\ln U_* \stackrel{\mr{SR}}{\approx} 52 - \ln \frac{0.07}{r} \ ,
\end{split}
\end{equation}
where $\Delta N_\reh$ is the number of e-folds between the end of inflation and the end of reheating, taken to be $\Delta N_\reh=4$ for SM field content \cite{Figueroa:2009jw,Figueroa:2015rqa,Repond:2016sol} (see also \cite{Ema:2016}), and $U_*$ is the potential at the Hubble exit of $k_*$. The last equality applies in the SR approximation.

\subsection{Special cases} \label{sec:inflation_cases}

\subsubsection{Slow-roll inflation} \label{sec:slowroll}

Let us define the second Hubble SR parameter:
\begin{equation} \label{eq:second_hubble_sr_parameter}
	\eta_H \equiv -\frac{\ddot{\chi}}{H\dot{\chi}} \ .
\end{equation}
If both $|\eta_H|$ and $\epsilon_H$ are $<1$, we have SR inflation. In this case the second equation in \eqref{eq:friedmann_scalar_eq} also simplifies, and we have
\begin{equation} \label{eq:SR_condition}
	|\ddot{\chi}| \ll 3 H |\dot{\chi}| \qquad \Rightarrow \qquad 3 H \dot{\chi} \approx -U'(\chi) \ .
\end{equation}
We can also define the potential SR parameters:
\begin{equation}
	\label{eq:pot_sr_parameters}
	\epsilon_V \equiv \frac{1}{2} \qty(\frac{U'}{U})^2 \ , \qquad
	\eta_V \equiv \frac{U''}{U}
\end{equation}
with
\begin{equation}
	\label{eq:sr_params_conversion}
	\epsilon_H \approx \epsilon_V \ , \qquad
	\eta_H \approx \eta_V - \epsilon_V \qquad \text{in SR.}
\end{equation}
Smallness of $\epsilon_V$ and $|\eta_V|$ is a necessary (but not sufficient) condition for SR. SR predicts the CMB observables \cite{Lyth:2009zz}
\begin{equation} \label{eq:sr_observables}
	n_s = 1-6\epsilon_V+2\eta_V \ , \qquad \ r = 16 \epsilon_V \ .
\end{equation}
SR inflation is desirable on CMB scales as it predicts a nearly scale invariant spectrum, $n_s \approx 1$, in agreement with the observed value \eqref{eq:ns_obs}.

\subsubsection{Ultra-slow-roll inflation} \label{sec:USR}

If the first derivative of the potential is negligible at some point during inflation, the equations of motion become
\begin{equation}
	\label{eq:friedmann_eq_USR}
	3 H^2 \approx U(\chi) \approx \text{constant} \ , \qquad
	\ddot{\chi} + 3H\dot{\chi} \approx 0 \ .
\end{equation}
This regime is called ultra-slow-roll inflation (USR) \cite{Faraoni:2000vg, Kinney:2005vj, Martin:2012pe, Dimopoulos:2017ged, Pattison:2018}. The SR approximation does not hold, since $\eta_H=3>1$, even though $\epsilon_V\approx0$ and we can have $|\eta_V|<1$. The solution of \eqref{eq:friedmann_eq_USR} is
\begin{equation} \label{eq:usr_solution}
	\dot{\chi}(t) \propto e^{-3Ht} \quad \Rightarrow \quad \chi(t) \propto A + e^{-3Ht} \ ,
\end{equation} 
where $A$ is a constant, so the field slows down exponentially. As we will see, the power spectrum $\PR$ is greatly enhanced during USR.

\subsubsection{Inflation near a critical point} \label{sec:critical}

Let us consider inflationary dynamics in the case when the inflaton potential has a critical point at $\chi_0$, that is, $U'(\chi_0)=U''(\chi_0)=0$ (see figure \ref{fig:inflection_point_sketch}). We further assume that $\chi$ starts in SR above this feature, rolling towards $\chi_0$. There are three possible outcomes, with different background evolution and perturbation power spectra, depending on the details of the potential.

\begin{enumerate}

\item If the SR approximation holds all the way to $\chi_0$, then it takes an infinite number of e-folds to reach the critical point. This case is not physically relevant, since $\chi$ never gets to the vacuum at $\chi<\chi_0$ (the field velocity vanishes as $U'=0$, as \eqref{eq:SR_condition} shows). However, we may consider a near-critical point with $\eta_V=0$, $0 < \epsilon_V \ll 1$, so that the field rolls over the feature in a finite time. As we will see below, the scalar power spectrum is enhanced in this case, but the SR approximation holds all the way through the feature.

\item If the SR approximation breaks down before the critical point, but inflation continues until $\chi < \chi_0$ --- that is, $\epsilon_V<1$ all the way through, but $|\eta_V - \epsilon_V| >1$ at least somewhere above $\chi_0$, see \eqref{eq:sr_params_conversion} --- then near the critical point the field is in USR \cite{Germani:2017bcs, Motohashi:2017kbs, Gong:2017qlj, Kannike:2017bxn, Ballesteros:2017fsr, Pattison:2017mbe}. The USR conditions \eqref{eq:friedmann_eq_USR} are satisfied for only a short span of field values near the critical point --- in our numerical solutions for Higgs inflation, the USR period typically lasts only for a few e-folds, and the transition in and out of USR is not sharp. However, during this period, the scalar power spectrum can be strongly enhanced. This growth can be further enhanced if we consider a near-critical point instead, with $\epsilon_V=0$ but $\eta_V > 0$ so that the potential has a local minimum. After the feature the field may return to SR or inflation may end, depending on the shape of the potential.

\item If $\epsilon_V>1$ at some point above but close to the critical point, then inflation ends before the critical point is reached, or at least leaves the SR attractor. Typically the inflaton rolls over the critical point quickly, and there is hardly any enhancement of the scalar power spectrum \cite{Germani:2017bcs}.

\end{enumerate}
The second case is particularly interesting:  the SR approximation fails in the middle of inflation, and the scalar power spectrum can be greatly enhanced. Below we look at the evolution of the scalar power spectrum in this case analytically, before doing a full numerical calculation.

\subsection{Perturbations} \label{sec:perturbations}

\subsubsection{Mode equation} \label{sec:mode_eq}

Scalar perturbations during inflation are typically described in terms of the Sasaki-Mukhanov variable $\nu$ \cite{Mukhanov:1988jd}. Its Fourier modes $\mu_k$ satisfy the equation of motion
\begin{equation} \label{eq:mode_eq}
	\mu_k'' + \qty(k^2 - \frac{z''}{z}) \mu_k = 0 \ ,
\end{equation}
where prime denotes derivative with respect to conformal time $\eta=\int dt/a(t)$, and \begin{equation} \label{eq:z}
	z \equiv a\frac{\dot{\chi}}{H}
\end{equation}
is determined by the background solution. We solve \eqref{eq:mode_eq} in the Bunch--Davies vacuum \cite{Birrell:1982ix}, corresponding to the initial conditions
\begin{equation} \label{eq:bunch_davies_mu}
	\mu'_k = -ik\mu_k \ , \qquad |\mu_k| = \frac{1}{\sqrt{2k}}
\end{equation}
at an early time when the mode is deep inside the Hubble radius, $aH \ll k$.

To compare to observations, we calculate from $\mu_k$ the power spectrum of the comoving curvature perturbation,
\begin{equation} \label{eq:PR_in_mu}
	\PR(k,t) = \frac{k^3}{2\pi^2}\frac{|\mu_k(t)|^2}{z(t)^2} \ .
\end{equation}
In SR inflation with adiabatic perturbations, $\PR(k,t)$ does not depend on $t$ after Hubble exit, $aH \gg k$. This is not true in USR, so we define
\begin{equation} \label{eq:PR_at_infinity}
	\PRfrozen(k) \equiv \lim_{t\to\infty} \PR(k,t)
\end{equation}
and compare this quantity to the observed value \eqref{eq:pert_norm} and its running or to the conditions for PBH formation.

Thus far, the treatment in this section has been exact in linear perturbation theory, with no assumptions about the background evolution. Next, we solve $\PR$ in some special cases, to get analytical understanding of its behaviour near a critical point.

\subsubsection{Slow-roll} \label{sec:SR_perts}

Let us first calculate $\PR$ in SR inflation to zeroth order in the SR parameters. In this approximation, $H$ and $\dot{\chi}$ are constant, $z(\eta) \propto a(\eta) \propto -\eta^{-1}$, and the solution of the mode equation \eqref{eq:mode_eq} with the initial conditions \eqref{eq:bunch_davies_mu} is
\begin{equation} \label{eq:mu_de_sitter}
	\mu_k = \frac{1}{\sqrt{2k}}\qty(1-\frac{i}{k\eta})e^{-ik\eta} \ .
\end{equation}
Inserting this and \eqref{eq:z} into \eqref{eq:PR_in_mu}, $\PR$ freezes on super-Hubble scales $-\eta k \ll 1$ to the scale-independent value
\begin{equation} \label{eq:PR_in_de_sitter}
	\PRfrozen = \frac{H^4}{4\pi^2 \dot{\chi}^2} \ .
\end{equation}
Calculating to first order in SR parameters modifies this result slightly by adding scale-dependence, but the super-Hubble freeze still happens. It is conventional to calculate $\PRfrozen(k)$ using an asymptotic expression evaluated at Hubble exit, giving
\begin{equation} \label{eq:PR_in_SR}
	\PRfrozen(k) \approx \frac{H^4}{4\pi^2 \dot{\chi}^2}\bigg\rvert_{k=aH}=\frac{H^2}{8\pi^2 \epsilon_H}\bigg\rvert_{k=aH}\approx\frac{V}{24\pi^2 \epsilon_V}\bigg\rvert_{k=aH} \ .
\end{equation}
This is the standard SR result for $\PRfrozen$. 

In the general, non-SR case it is convenient to define a new variable, especially for the numerical considerations of section \ref{sec:scans}:
\begin{equation} \label{eq:g}
	g_k \equiv \frac{\mu_k e^{ik\eta}}{z} \ .
\end{equation}
Multiplication by $e^{ik\eta}$ removes the rapid phase oscillations on sub-Hubble scales present in \eqref{eq:mu_de_sitter}, and the factor $1/z$ absorbs all of the time-dependence of $\PR(k,t)$ into $g_k(t)$:
\begin{equation} \label{eq:PR_in_g}
	\PR(k,t) = \frac{k^3}{2\pi^2}|g_k(t)|^2 \ .
\end{equation}
When $\PR$ freezes to a constant value at super-horizon scales, so does $g_k(t)$, unlike $\mu_k$ whose exponential growth obscures details of the solution.

In terms of $g_k$, the Bunch--Davies initial conditions read
\begin{equation} \label{eq:bunch_davies_g}
	\dot{g}_k = -\frac{\dot{z}}{z}g_k \ , \qquad |g_k| = \frac{1}{\sqrt{2k} |z|} \quad \mr{at} \quad k \gg aH \ ,
\end{equation}
and the mode equation \eqref{eq:mode_eq} becomes
\begin{equation} \label{eq:mode_eq_g}
	\ddot{g}_k + \qty(H + 2\frac{\dot{z}}{z} - \frac{2ik}{a})\dot{g}_k - \frac{2ik}{a}\frac{\dot{z}}{z}g_k = 0 \ ,
\end{equation}
where we have switched back to cosmic time $t$ for easier numerical treatment --- the dynamical range of $t$ is smaller than that of the exponentially changing $\eta$. Here
\begin{equation} \label{eq:zdotz}
	\frac{\dot{z}}{z} = H(1+\epsilon_H-\eta_H) \ .
\end{equation}
This is an exact expression. At zeroth order in SR, $\dot{z}/z=H$, and in super-Hubble limit $k\ll a H$, equation \eqref{eq:mode_eq_g} reduces to
\begin{equation} \label{eq:mode_eq_g_approx_SR}
	\ddot{g}_k + 3H\dot{g}_k = 0 \quad \Rightarrow \quad \dot{g}_k \propto e^{-3Ht} \quad \Rightarrow \quad g_k \propto A + e^{-3Ht} \ ,
\end{equation}
where $A$ is a constant. The mode $g_k$ and hence the power spectrum $\PR(k,t)$ freeze to constant values as time goes on, as discussed above. Taking $k$-dependence into account, two independent solutions of \eqref{eq:mode_eq_g} at zeroth order in SR are
\begin{equation} \label{eq:g_SR}
	g_{k1} = \qty(i+\frac{k}{aH})  \ , \qquad
	g_{k2} = \qty(i-\frac{k}{aH})e^{-2ik/(aH)} \ ,
\end{equation}
and the solution with the Bunch--Davies initial conditions \eqref{eq:bunch_davies_g} is
\begin{equation} \label{eq:g_de_sitter}
	g_k = \frac{1}{\sqrt{2k}}\frac{H^2}{k\dot{\chi}}g_{k1} \ .
\end{equation}

\subsubsection{Ultra-slow-roll} \label{sec:USR_perts}

In USR, we have $\eta_H\approx3$, $\epsilon_H\approx0$, so $\dot{z}/z\approx-2H$. The equation corresponding to \eqref{eq:mode_eq_g_approx_SR} reads
\begin{equation} \label{eq:mode_eq_g_approx_USR}
	\ddot{g}_k - 3H\dot{g}_k = 0 \quad \Rightarrow \quad \dot{g}_k \propto e^{3Ht} \quad \Rightarrow \quad g_k \propto A + e^{3Ht} \ ,
\end{equation}
where $A$ is a constant. We see that in USR, the spectrum does not freeze on super-Hubble scales, but is instead exponentially amplified. The mode function grows without limit as long as USR lasts. This is a known phenomenon \cite{Kinney:2005vj}; we explain in appendix \ref{app:PRgrowth} why the usual arguments about modes freezing in the super-Hubble limit fail.

Exact solutions of \eqref{eq:mode_eq_g} to zeroth order in USR, with constant $H$ and $\dot{z}/z=-2H$, are linear combinations of
\begin{equation} \label{eq:g_USR}
	g_{kA} = \qty(i+\frac{k}{aH})a^3 \ , \qquad
	g_{kB} = \qty(i-\frac{k}{aH})a^3e^{-2ik/(aH)} \ .
\end{equation}

\subsubsection{Tensor perturbations} \label{sec:tensor_perts}

At this point, let us mention how the story goes for the tensor perturbations. Their modes follow the equation \cite{Lyth:2009zz}
\begin{equation} \label{eq:mode_eq_h}
	h_k'' + \qty(k^2 - \frac{a''}{a})h_k = 0 \ .
\end{equation}
This is equal to \eqref{eq:mode_eq} with $z=a$, so $\dot{z}/z=H$ always, in both SR and USR. Thus the tensor equivalent of the power spectrum \eqref{eq:PR_in_mu} is not amplified, and the standard result
\begin{equation} \label{eq:PT_in_SR}
	\mathcal{P}_\mathcal{T}(k) = 8\qty(\frac{H}{2\pi})^2\bigg\rvert_{k=aH}
\end{equation}
applies as long as inflation lasts. Of course, the tensor-to-scalar ratio $r$ is modified if the scalar power spectrum changes.

\subsubsection{Critical point} \label{sec:critical_perts}

Let us then estimate the evolution of $\PR$ in the vicinity of a critical point in the case when the inflaton starts in SR, rolls over the critical point in USR, and then returns to SR. During the first SR period, the field velocity is approximately constant, $\dot{\chi}=\dot{\chi}_1$; during USR $|\dot{\chi}|$ decreases exponentially; and after USR ends, $\dot{\chi}$ stays constant at its new smaller value. Let $H$ be approximately constant all the way through. Then $\dot{z}/z$ \eqref{eq:zdotz} jumps from $H$ to $-2H$ instantly at time $t_1$ when USR starts, and jumps back to $H$ at time $t_2$ when USR ends.

\begin{figure}
\begin{center}
\includegraphics{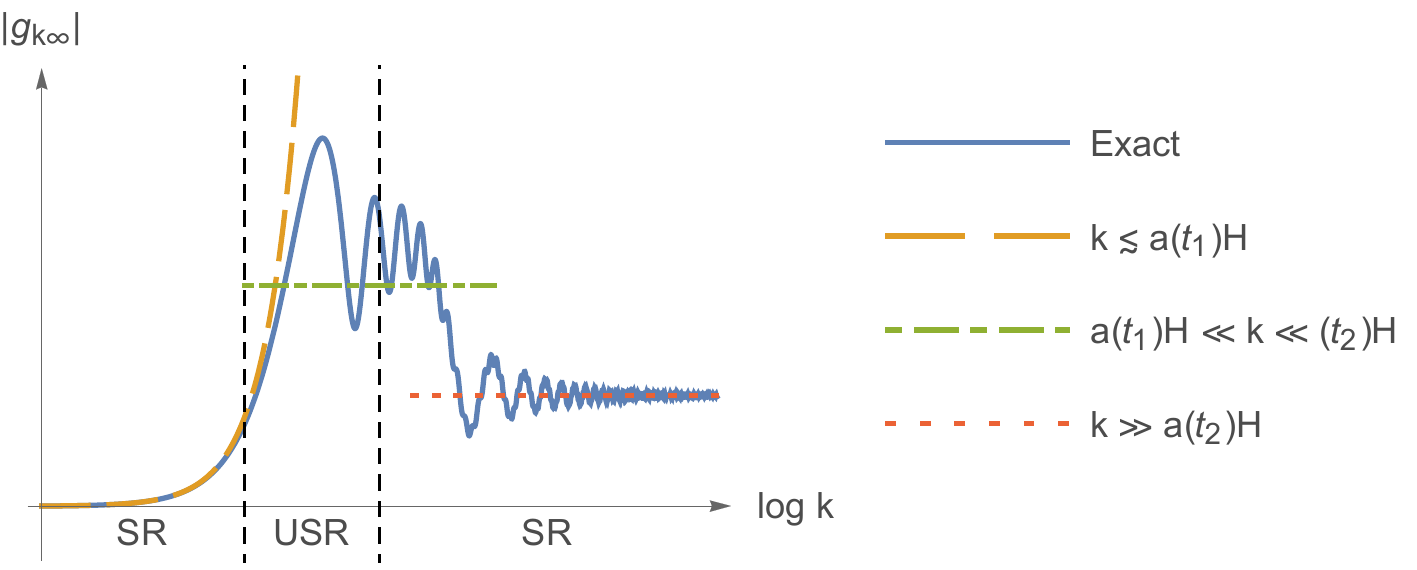}
\end{center}
\caption{The final frozen amplitude of the mode function $g_{k\infty}$ in the critical point model of section \ref{sec:critical_perts}. We show the exact solution (solid line) and the USR approximation \re{eq:gk_inflection_point_approx} for different $k$-values (dashed lines). Modes on the left (small $k$) exit the Hubble radius during the initial SR period; modes in the middle exit the Hubble radius during USR; and modes on the right (large $k$) exit the Hubble radius during the final SR period. The oscillatory behaviour is an artefact due to the instant jumps between SR and USR. It is absent in the numerical calculations where the transitions are smooth, see figures \ref{fig:PR_in_N} and \ref{fig:PR_in_lnk_peak}.}
\label{fig:gk_analytical}
\end{figure}

For $t<t_1$, the mode function $g_k$ follows the solution \eqref{eq:g_de_sitter} and approaches the constant SR value $G \equiv iH^2/(2k^{3/2}\dot{\chi}_1)$. After $t_1$, $g_k$ is a linear combination of the solutions \eqref{eq:g_USR}, and after $t_2$ it is a linear combination of the solutions \eqref{eq:g_SR}. We solve $g_k(t)$ by matching these solutions and their first derivatives at the transition times. Since the field ends in SR, the mode function freezes to a constant value at $t \gg t_2$ that depends on the time of Hubble exit. The leading behaviour is
\begin{equation} \label{eq:gk_inflection_point_approx}
	g_{k\infty} \equiv \lim_{t\to\infty} g_k(t) =
	\begin{cases}
      G & \qquad k \ll a(t_1)H \\
      G\qty(1-\frac{4}{5}e^{3N_\mr{USR}-2N_1}) & \qquad k \lesssim a(t_1)H \\
      2Ge^{3N_\mr{USR}} & \qquad a(t_1)H \ll k \ll a(t_2)H \\
      Ge^{3N_\mr{USR}} & \qquad k \gg a(t_2)H \ ,
   \end{cases}
\end{equation}
where $N_\mr{USR}=(t_2-t_1)H$ is the number of e-folds of USR, assumed to be large, and $N_1$ is the number of e-folds between Hubble exit and the start of USR, assumed to be small. Modes that exit the Hubble radius deep in the initial SR phase are not affected by USR, whereas modes that exit near the beginning of or during USR are amplified. For $N_1 \gg \frac{3}{2} N_\mr{USR}$ there is no amplification: the mode function has already settled to the value $G$ before the beginning of USR and doesn't significantly change afterwards. Amplification starts for $N_1 \lesssim \frac{3}{2}N_\mr{USR}$. The mode function $g_k$ is amplified by a factor proportional to $e^{3N_\mr{USR}}$, so $\PRfrozen$ is amplified proportional to $e^{6N_\mr{USR}}$. The maximum amplification occurs for the USR scales, with a lower plateau afterwards. An exact solution for $g_k$ together with the above approximations is plotted in figure \ref{fig:gk_analytical}.

The result \eqref{eq:gk_inflection_point_approx} can be compared to the SR approximation for $g_k$ corresponding to \eqref{eq:PR_in_SR}:
\begin{equation} \label{eq:gk_inflection_point_SR}
	g_{k\infty} =
	\begin{cases}
      G & \qquad k \leq a(t_1)H \\
      G\qty(\frac{k}{a(t_1)H})^3 & \qquad a(t_1)H \leq k \leq a(t_2)H \\
      Ge^{3N_\mr{USR}} & \qquad k \geq a(t_2)H \ .
   \end{cases}
\end{equation}
This agrees with the USR approximation for $k \leq a(t_1)H$ and $k \geq a(t_2)H$, but rises monotonically in-between, while our result \eqref{eq:gk_inflection_point_approx} shows that there is an extra enhancement factor of two during USR. Note that the last expression in \eqref{eq:PR_in_SR} diverges during USR when $\epsilon_V=0$.

The model presented here is just a crude approximation. To get accurate results for Higgs inflation, we solve the mode equation \eqref{eq:mode_eq_g} with the initial conditions \eqref{eq:bunch_davies_g} numerically. Nevertheless, these examples demonstrate the effect of USR on the power spectrum: scales that exit the Hubble radius during USR are amplified without limit as long as USR lasts, and scales that exit just before or after USR are also amplified. This gives a peak in the power spectrum around the USR scales, not captured by the SR approximation \eqref{eq:PR_in_SR}, as shown in figure \ref{fig:gk_analytical}. If the peak is high enough, it can lead to production of primordial black holes after inflation when the scales re-enter the Hubble radius.

\section{Primordial black holes} \label{sec:PBHs}

\subsection{PBH mass} \label{sec:PBH_masses}

When modes with large amplitude re-enter the Hubble radius after inflation, a particularly dense Hubble patch may collapse into a black hole. Mass of inflationary PBHs formed from perturbations on scale $k$ is
\begin{equation} \label{eq:bhmass}
	M_\mr{PBH} = \gamma \frac{4\pi}{3}R^3\rho \ ,
\end{equation}
where $R=H^{-1}=a/k$ is the Hubble radius, $\rho$ is the total energy density at the time of Hubble entry of $k$, and $\gamma$ is an efficiency factor. We use the value $\gamma=0.2$ estimated to apply during radiation domination \cite{Carr:1975qj}. Accretion and other effects could significantly change $\gamma$ up or down, but it is straightforward to scale the masses correspondingly, and our results are robust to changes in $\gamma$ of several orders of magnitude. In the radiation-dominated era, the scale factor, energy density and Hubble parameter scale as
\begin{equation} \label{eq:radiationdomination}
	a(t) \propto t^{1/2} \ , \quad \rho(t) \propto a(t)^{-4} \ , \quad H(t) = \frac{1}{2t} \propto a(t)^{-2} \ .
\end{equation}
Thus, in terms of the wavenumber $k=aH$, we have
\begin{equation} \label{eq:bhmass_in_k}
	M_\mr{PBH} \propto k^{-2} \ .
\end{equation}
We can also relate $k$ to the number of e-folds during inflation. We must have $k>k_*$, where $k_*=0.05$ Mpc$^{-1}$ is the pivot scale, otherwise the PBHs will form too late and be too massive to constitute dark matter. So the PBHs form before the pivot scale re-enters, and the number of e-folds of inflation from the Hubble exit of scale $k_*$ to the Hubble exit of scale $k$ is
\begin{equation} \label{eq:bhmass_N}
	\Delta N \equiv N_* - N = \log \frac{a}{a_*} \approx \log \frac{aH}{a_*H_*} = \log \frac{k}{k_*} > 0 \quad \Rightarrow \quad M_\mr{PBH} \propto e^{-2 \Delta N} \ .
\end{equation}
We approximate that \eqref{eq:radiationdomination} holds all the way to matter-radiation equality and insert the values $M_\mr{eq}\approx6\gamma\times10^{50}$ g, $k_\mr{eq} \approx 0.01$~Mpc$^{-1}$ (using $\omega_\mr{m}=0.14$ \cite{Akrami:2018odb}) to normalise the mass of PBHs formed on scale $k$ as
\begin{equation} \label{eq:bhmass_N2}
	M_\mr{PBH} = \frac{M_\mr{eq}}{25}e^{-2\Delta N} \ .
\end{equation}
This equation gives PBH mass in a form that is easy to incorporate in inflationary analysis.

\subsection{PBH energy density fraction} \label{sec:PBH_massfraction}

We can also ask which fraction of the mass of the universe ends up in PBHs. Assuming Gaussian statistics, the probability distribution of perturbations at scale $k$ is \cite{Carr:1975qj, Green:2004wb, Young:2014ana}
\begin{equation} \label{eq:gaussiandistribution}
	P(\zeta_k) = \frac{1}{\sqrt{2\pi}\sigma_k} e^{-\frac{\zeta_k^2}{2\sigma^2_k}} \ ,
\end{equation}
where we approximate the width as $\sigma^2_k\approx\PRfrozen(k)$. A more accurate treatment with a window function \cite{Ando:2018} can lead to an order-of magnitude change in the value of $\sigma^2_k$. It is assumed that a PBH forms out of every region where $\zeta_k$ exceeds a threshold value $\zeta_c$, which has been estimated as $\zeta_c=0.07\dots1.3$ \cite{Carr:1975qj, Niemeyer:1999ak,Musco:2004ak,Harada:2013epa, Young:2014ana, Musco:2012au, Motohashi:2017kbs}. Assuming PBHs to instead form on peaks of the density field may be more appropriate, and this may affect the PBH abundance and mass \cite{Green:2004wb, Young:2014ana, Yoo:2018, Germani:2018jgr}. Neither this nor the change in $\sigma^2_k$ would change our conclusions regarding Planck scale PBHs, as we discuss in section \ref{sec:discussion}. The initial fraction of the energy density of the universe which ends up in these PBHs is then
\begin{equation} \label{eq:pbhfraction}
	\beta_k = 2 \int_{\zeta_c}^{\infty} P(\zeta_k) d\zeta_k = \mr{erfc}\qty(\frac{\zeta_c}{\sqrt{2 \PRfrozen(k)}}) \approx \frac{\sqrt{2 \PRfrozen(k)}}{\sqrt{\pi}\zeta_c} e^{-\frac{\zeta_c^2}{2 \PRfrozen(k)}} \ ,
\end{equation}
where erfc is the complementary error function, and the last approximation holds for $\zeta_c \gg \sqrt{\PRfrozen(k)}$.

After the PBHs have formed, their energy density scales like cold matter, $\rho_\mr{PBH} \propto a^{-3}$, whereas the energy density of radiation is diluted faster, $\rho_\mr{rad} \propto a^{-4}$. Thus during radiation domination, the PBH energy density fraction grows as $\rho_\mr{PBH}/\rho_\mr{rad} \propto a \propto k^{-1}$. In addition, if $\PRfrozen$ is enhanced over a wide range of scales, PBHs form on different scales and with different masses, and we have to sum over their contributions to get the total PBH energy density fraction. However, we will find that in critical Higgs inflation $\PRfrozen$ is peaked at a single scale, so the PBH spectrum is monochromatic to a good approximation.

Taking these considerations into account, the fraction of energy density in PBHs at matter-radiation equality is approximately
\begin{equation} \label{eq:omegaPBH}
	\Omega_\mr{PBH\,eq} = \frac{k}{k_\mr{eq}}\beta_{k} = 5 \frac{k}{k_*}\mr{erfc}\qty(\frac{\zeta_c}{\sqrt{2 \PRfrozen(k)}}) \approx 5\frac{\sqrt{2\PRfrozen(k)}}{\sqrt{\pi}\zeta_c}e^{-\frac{\zeta_c^2}{2\PRfrozen} + \Delta N} \ .
\end{equation}
The exponent is the determining factor. The maximum value of the power spectrum $\PRfrozen$ must be big enough so that the two terms in the exponent are of same magnitude, or $\Omega_\mr{PBH\,eq}$ will be exponentially small. Even for the minimum value $\zeta_c=0.07$, we need $\PRfrozen$ to be at least of order $10^{-4}$ to compensate for a few dozen e-folds in $\Delta N$. On the other hand, for $\PRfrozen\gtrsim1$, black holes are overproduced: the universe becomes matter dominated too early, and the fraction of matter in the dark sector is too large. The exact value of $\PRfrozen$ needed depends on $\Delta N$ and is sensitive to uncertainties in the above analysis.

\subsection{Planck mass relics} \label{sec:relics}

The above analysis applies for classical black holes. When quantum physics is taken into account, primordial black holes evaporate due to Hawking radiation. There are stringent constraints on evaporating PBHs since the evaporation could, for example, spoil BBN or produce too many gamma ray bursts \cite{Carr:2009jm, Carr:2016drx, Carr:2017jsz}. However, there are no constraints for black holes that evaporate before the EW crossover. If the evaporation is not complete, but Planck mass relics are left behind, they could constitute dark matter \cite{MacGibbon:1987my, Barrow:1992hq, Carr:1994ar, Green:1997sz, Alexeyev:2002tg, Chen:2002tu, Barrau:2003xp, Chen:2004ft, Nozari:2005ah}.

For PBHs to evaporate early enough and not spoil baryogenesis, their mass has to be less than $10^6$ g (see appendix \ref{app:uppermasslimit}). Using \eqref{eq:bhmass_N2}, this corresponds to $\Delta N \gtrsim 49$, that is, the perturbations that seed the PBHs must exit the Hubble radius near the end of inflation (in Higgs inflation typically $N\approx50$, as \eqref{eq:Npivot} shows).

If the relics have mass $M_\mr{rel}$, their fraction of the energy density at matter-radiation equality is approximately
\begin{equation} \label{eq:relic_omega}
	\Omega_\mr{rel\,eq} = \frac{M_\mr{rel}}{M_\mr{PBH}}\frac{k}{k_\mr{eq}}\beta_{k} \approx \frac{M_\mr{rel}}{\Mpl}\beta_{k} e^{3\Delta N - 123} \ ,
\end{equation}
where $\Mpl=1/\sqrt{8\pi G_\mr{N}}$ is the Planck mass, and we used \eqref{eq:bhmass_N2} for $M_\mr{PBH}$. For $M_\mr{rel}=\Mpl$, $\Delta N = 50$, this is of order unity for $\beta_{k} \sim e^{-30}$, which according to \eqref{eq:pbhfraction} corresponds to $\PRfrozen\approx10^{-2}\zeta_c^2 \sim 10^{-4}\dots 10^{-2}$. If $\PRfrozen$ can be enhanced to this magnitude near the end of inflation, the relics can constitute part or all of dark matter.

\section{Numerical analysis} \label{sec:scans}

\subsection{Scan method}

To study PBH production in Higgs inflation, we construct the Higgs potential with quantum corrections and renormalization group equation running as discussed is section \ref{sec:higgs_pot}, scan over possible critical point and near-critical point Higgs potentials, and calculate the power spectra $\PRfrozen(k)$ by solving the mode equations \eqref{eq:mode_eq_g} numerically.

We have four parameters: the non-minimal coupling $\xi$, the location of the feature given by $\delta_0$ defined in \eqref{eq:delta} (the subscript 0 denotes quantities evaluated at the feature) and the jumps $\Delta\lambda$ and $\Delta y_t$ in the Higgs quartic coupling and top Yukawa coupling, respectively. In practice, we swap $\Delta\lambda$ and $\Delta y_t$ for the slow roll parameters $\epsilon_{V0}$ and $\eta_{V0}$. Choosing these four numbers fixes the potential and its quantum corrections. An exact critical point corresponds to $\epsilon_{V0}=\eta_{V0}=0$, but we allow for small deviations around this. The allowed values of $\delta_0$ vary around unity. The initial values for couplings $\lambda$ and $y_t$, as well as the constant $\kappa$ in \eqref{eq:jumpscale}, are determined from the conditions that the feature is formed at the given scale and the one-loop correction \eqref{eq:one_loop_corr} vanishes there. The potential is then determined as described in section \ref{sec:higgs_pot}. Afterwards, the values of $\lambda$ and $y_t$ run down from the feature are compared to their SM counterparts run up from the EW scale to determine the jumps $\Delta \lambda$ and $\Delta y_t$ needed at the threshold scale \eqref{eq:jumpscale}. We use the mean values \eqref{eq:SM_bestfit_vals} for the values of the SM parameters at the EW scale. We discuss the dependence on these EW scale values in section \ref{sec:jumps}.

After fixing the potential, we solve the background equations \eqref{eq:friedmann_scalar_eq} numerically. Then we calculate the frozen super-Hubble values $\PRfrozen$ by using the approximation \eqref{eq:PR_in_SR} during SR and solving the mode equations \eqref{eq:mode_eq_g} numerically outside SR. We find the maximum of $\PRfrozen$ near the feature scale and compare it to the threshold value $10^{-4}$ needed for PBH production.

\begin{figure}
\begin{center}
\includegraphics{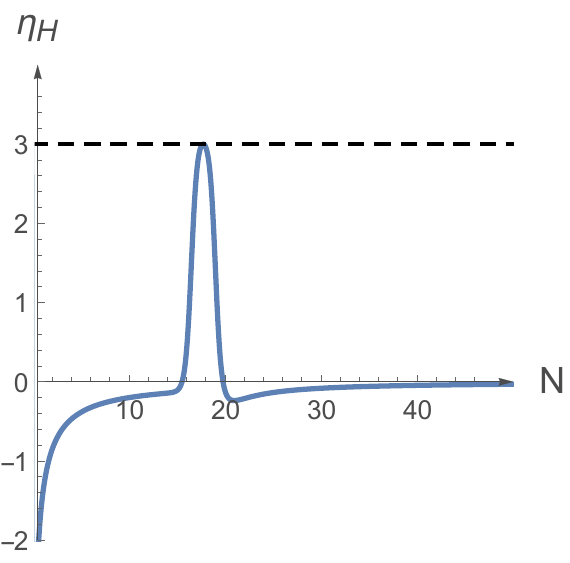}
\end{center}
\caption{The SR parameter $\eta_H$ as a function of the e-folds $N$ for the potential of figure \ref{fig:inflection_point_sketch}. The parameter $\eta_H$ briefly goes up to 3 during a USR period around $N=18$.}
\label{fig:etaH_in_N}
\end{figure}

\begin{figure}
\begin{center}
\includegraphics{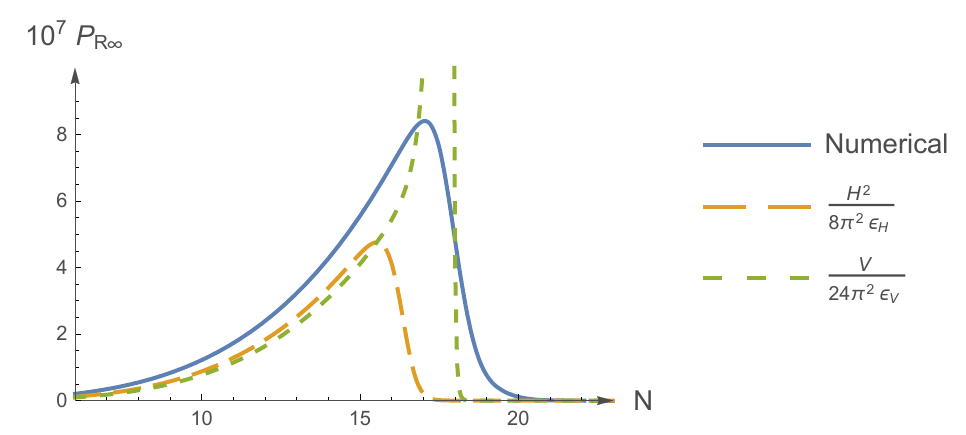}
\end{center}
\caption{The power spectrum $\PRfrozen$ as a function of $N$ at the Hubble exit of the corresponding mode (solid line) and SR approximations from \eqref{eq:PR_in_SR} (dashed lines) for the case of figures \ref{fig:inflection_point_sketch} and \ref{fig:etaH_in_N}. The power spectrum has a peak at the USR scale, and the SR approximations fail, although the peak in the $\epsilon_H$ approximation has the right order of magnitude.}
\label{fig:PR_in_N}
\end{figure}

An example of this procedure is sketched in figures \ref{fig:etaH_in_N} and \ref{fig:PR_in_N}, where we use the potential shown in figure \ref{fig:inflection_point_sketch}. For these parameter values, there is a USR period around $N\approx 18$. The SR approximation for $\PRfrozen$ is clearly inadequate to describe $\PRfrozen$ at the USR scales.

In addition to probing the PBH scales, the model has to match CMB observations. We demand that the amplitude $\PRfrozen(k_*)$, spectral index $n_s$, and the tensor-to-scalar ratio $r$ at the pivot scale match the observed values \eqref{eq:pert_norm}--\eqref{eq:r_obs}. Scanning over the free parameters with these restrictions and $\epsilon_{V0}=\eta_{V0}=0$ doesn't produce high enough peaks in $\PRfrozen$ for PBH formation. We then relax the exact critical point condition and let $\epsilon_{V0}$ and $\eta_{V0}$ vary. There are two distinct possibilities, corresponding to cases 1 and 2 in section \ref{sec:critical}. First, if the field is in SR all the way through the feature, we have to choose $\epsilon_{V0}>0$ so that the field rolls over the feature in a finite time. This case is relevant for features at $\delta_0 \lesssim 1$. Second, if there is a USR period near the feature, we can choose $\eta_{V0} > 0$ to enhance the peak in the power spectrum. In the metric formulation, this happens for $\delta_0 \gtrsim 1$, while in the Palatini formulation we still have $\delta_0<1$. Let us consider these cases separately.

\subsection{SR near the feature} \label{sec:scan_d0_small}

If the feature is formed at a field value that at tree-level corresponds to the exponentially flat plateau, the SR parameters are small and the field is in SR all the way through the feature. We can then use the SR result \eqref{eq:PR_in_SR} to calculate $\PRfrozen$. To get over the feature in a finite number of e-folds, we consider a near-critical point with $\eta_{V0}=0$ but $\epsilon_{V0}>0$. We have scanned over parameters $\delta_0$ and $\xi$, and fixed $\epsilon_{V0}$ so that the amplitude of scalar perturbations agrees with the CMB observations \eqref{eq:pert_norm} at the pivot scale, restricting the scan to cases where the SR approximation holds until the end of inflation. The peak amplitude of $\PRfrozen$ turns out to be below $10^{-4}$ everywhere. Varying also $\eta_{V0}$ around zero does not change order of magnitude of the peak values of $\PRfrozen$. It is thus impossible to produce a large amount of PBHs from SR while also matching the CMB observations. Similar conclusions apply in the Palatini formulation.

\subsection{USR near the feature} \label{sec:scan_d0_big}

\subsubsection{Critical point or a local minimum}

\begin{figure}
\begin{center}
\includegraphics{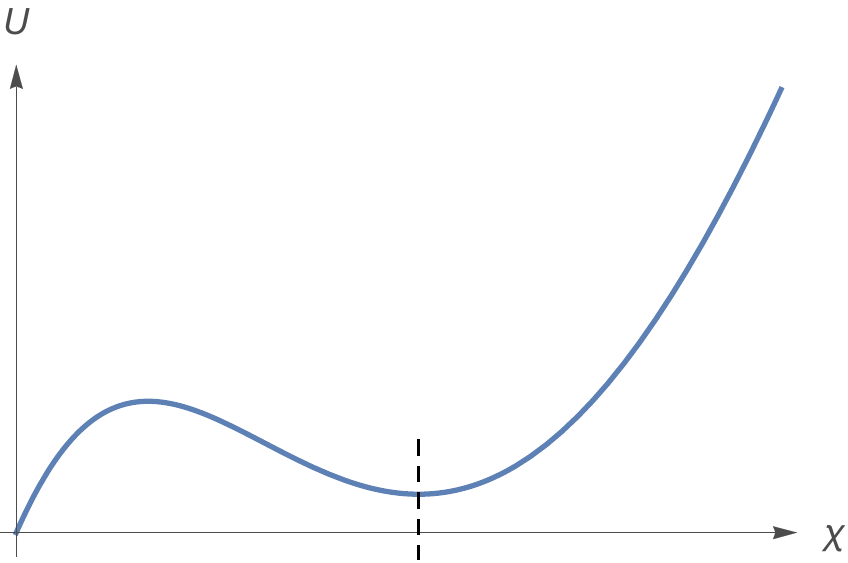}
\end{center}
\caption{Sketch of a potential with a shallow local minimum. Dashed line marks $\delta_0$, where $\epsilon_V=0$, $\eta_V>0$. This potential was formed with parameter values $\xi=80$, $\delta_0=1.53$, $\epsilon_{V0}=0$, and $\eta_{V0}=15.037$ in the metric formulation.}
\label{fig:potential_with_minimum}
\end{figure}

If the feature is formed just below the plateau scales, the field is in USR near the feature. In this case the SR parameters, in particular $\eta_V$, become large before the feature is reached and the SR approximation fails. We solve the mode equations \eqref{eq:mode_eq_g} numerically to calculate $\PRfrozen$, and scan again over $\delta_0$ and $\xi$, now demanding that SR is broken before the end of inflation. With a pure critical point, $\epsilon_{V0}=\eta_{V0}=0$, not many PBHs are formed: for all such potentials compatible with CMB observations, we get $\PRfrozen \lesssim 10^{-6}$. We thus let $\eta_{V0}$ take positive values while fixing $\epsilon_{V0}=0$. As a result, the potential has a shallow local minimum instead of an exact critical point, see figure \ref{fig:potential_with_minimum}. If the minimum is too deep, the inflaton will get stuck, but by fine-tuning $\eta_{V0}$ we can make the inflaton roll through the feature in just the right amount of e-folds.

\begin{figure}
\begin{center}
\includegraphics{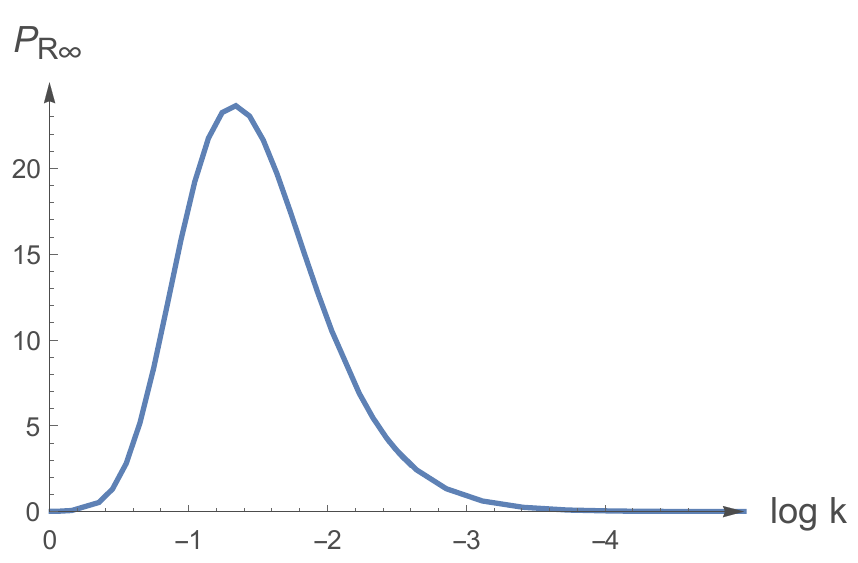}
\end{center}
\caption{The power spectrum $\PRfrozen$ for the potential of figure \ref{fig:potential_with_minimum} with a local minimum. Inflation stops briefly as the inflaton passes the feature, so it is more convenient to use $\log k$ instead of $N$ on the $x$-axis, normalized so that $\log k=0$ for the scale whose Hubble exit happens at the end of inflation. The SR expression for $\PRfrozen$ with $\epsilon_H$ \eqref{eq:PR_in_SR} is a poor approximation for this potential: it has a maximum magnitude of $10^{-6}$. In this particular case, the peak in $\PRfrozen$ is formed near the end of inflation and is narrower than in figure \ref{fig:PR_in_N}, since there is no subsequent SR period after USR.}
\label{fig:PR_in_lnk_peak}
\end{figure}

As pointed out in \cite{Kannike:2017bxn, Ballesteros:2017fsr, Bezrukov:2017dyv}, a shallow minimum can further enhance $\PRfrozen$. The inflaton decelerates rapidly when climbing up a hill, which increases the parameter $\eta_H$ \eqref{eq:second_hubble_sr_parameter} --- even beyond the usual USR value of 3 --- and amplifies the power spectrum, as discussed in section \ref{sec:perturbations}. An example of $\PRfrozen$ in such a situation is plotted in figure \ref{fig:PR_in_lnk_peak}. The SR approximation \eqref{eq:PR_in_SR} is even less valid here than in the critical point case shown in figure \ref{fig:PR_in_N}; the value of $\PRfrozen$ can be orders of magnitude above the SR approximation. In particular, $\PRfrozen$ can be above one even when the SR approximation \eqref{eq:PR_in_SR} is below the threshold $10^{-4}$. Hence a numerical calculation of $\PRfrozen$ is crucial.

\begin{figure}
\begin{center}
\includegraphics{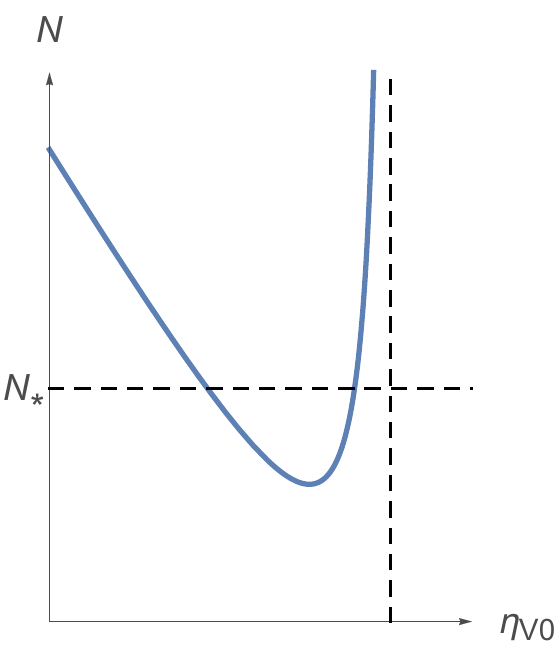}
\end{center}
\caption{Sketch of the relation between $N$ and $\eta_{V0}$. Usually $N$ approaches infinity for a finite positive $\eta_{V0}$; however, when the potential at the bottom of the local minimum approaches zero, $N$ becomes bounded from above. There may be zero, one or two $\eta_{V0}$-values corresponding to the wanted $N=N_*$, depending on the height of the minimum of the curve.}
\label{fig:eta0_in_N}
\end{figure}

It turns out that for some values of $\delta_0$ and $\xi$ there are two possible values of $\eta_{V0}$ that give a pivot scale spectrum in agreement with observations. To understand why, let's denote the number of e-folds of inflation corresponding to the scale with the right amplitude of scalar perturbations by $N$, and check how $N$ changes as $\eta_{V0}$ is changed. When $\eta_{V0}$ increases, so does the value of $\eta_V$ near the pivot scale. This makes the breaking of the SR approximation stronger and makes the field `overshoot' the feature more: the inflaton field reaches a higher velocity before the feature and rolls over it in a shorter time. This effect decreases $N$. However, increasing $\eta_{V0}$ also makes the local minimum deeper, so it takes more time for the inflaton to climb up to the other side. This increases $N$. These two effects compete, and in practice $N$ depends on $\eta_{V0}$ as shown in figure \ref{fig:eta0_in_N}: as $\eta_{V0}$ increases from zero, $N$ first decreases and then increases, approaching infinity at some $\eta_{V0}$ when the minimum becomes so deep that the inflaton gets stuck. However, the requirement that the potential stays positive sets a maximum depth for the minimum: if a deeper minimum would be needed to get a big enough $N$, then the correct $N$ can't be achieved. The right $N$ can then be reached for either two values of $\eta_{V0}$ (`high' and `low' $\eta_{V0}$), one value of $\eta_{V0}$ (corresponding to the `high' case), or no values of $\eta_{V0}$ at all; see figure \ref{fig:eta0_in_N}.

Note that in a potential with a local minimum, we always have $\epsilon_{V0}=0$ and $\eta_{V0}>0$ at the minimum. As the peak in $\PRfrozen$ would not be high enough without the minimum, our scan with $\epsilon_{V0}=0$ and $\eta_{V0}>0$ covers all potentials that are interesting for PBH formation from USR.

\subsubsection{Results} \label{sec:results_d0_big}

\begin{figure}
\begin{center}
\begin{tabular}{ll}
\includegraphics{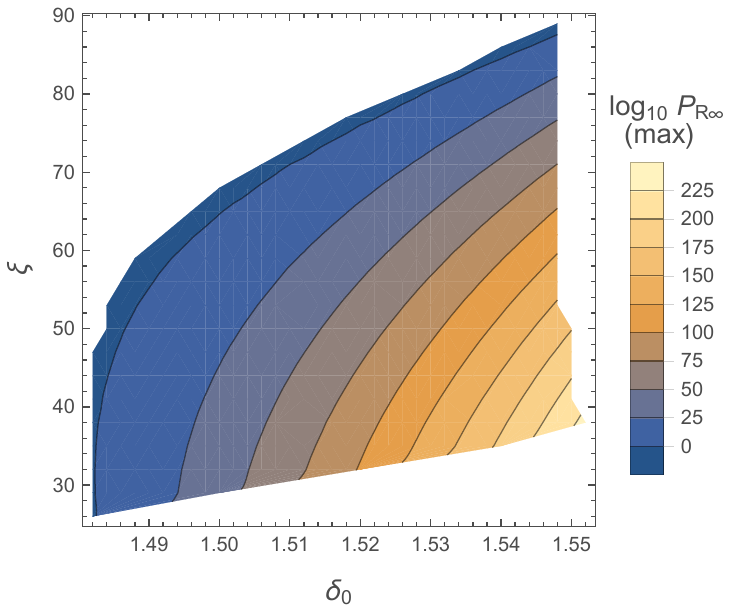} &
\includegraphics{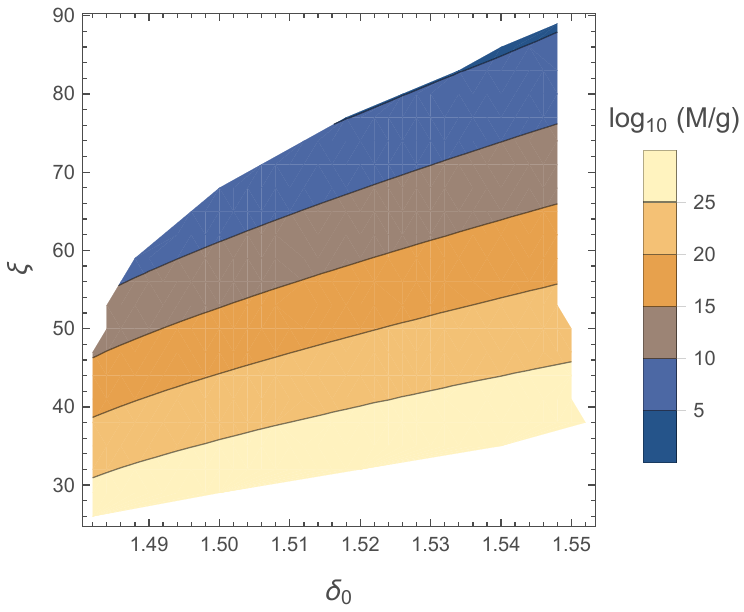} \\
\includegraphics{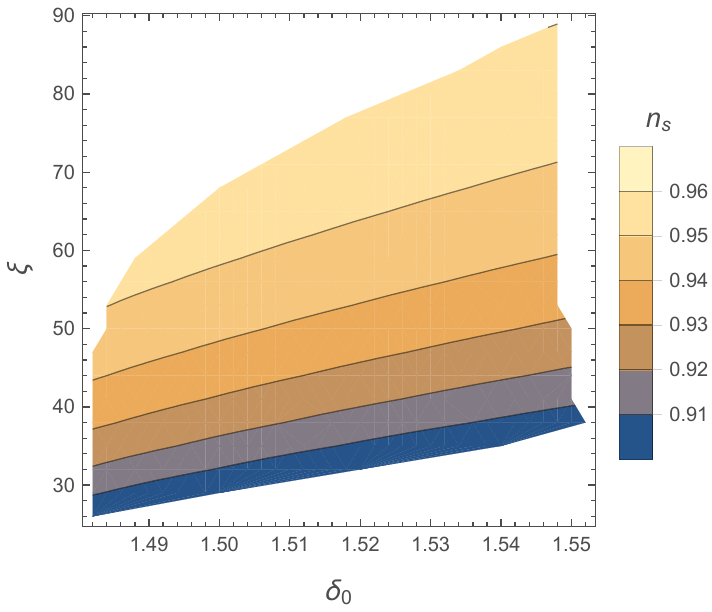} &
\includegraphics{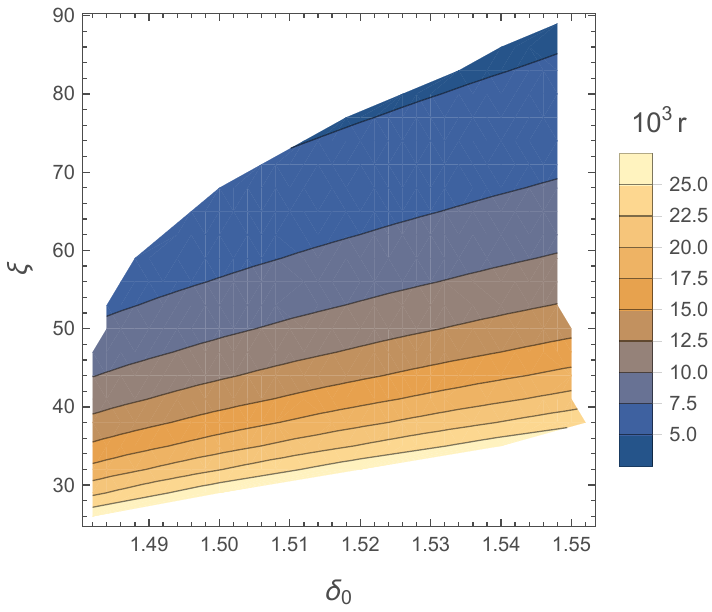}
\end{tabular}
\end{center}
\caption{
Results of the USR scan on the high $\eta_{V0}$ branch: peak value of $\PRfrozen$ (top left), PBH mass in grams (top right), spectral index $n_s$ (bottom left) and tensor-to-scalar ratio $r$ (bottom right) in the metric formulation. These figures come from the metric case. The corresponding figures in the Palatini case are similar, but the exact numbers are different, in particular the ranges of $\delta_0$ and $\xi$ --- see the text.}
\label{fig:scans_big_d0_high_eta0}
\end{figure}

For the low $\eta_{V0}$ branch we always have $\PRfrozen<10^{-4}$, so the number of PBHs is too small. However, the high $\eta_{V0}$ branch can produce PBHs in agreement with observations. Results of the numerical scan are shown in figure \ref{fig:scans_big_d0_high_eta0} for the metric case. We see that by tuning the values of $\xi$ and $\delta_0$ we can produce as many PBHs as desired, and with any desired mass up to at least $M \sim 10^{35}$ g. In particular, after fixing all other parameters from observations, by changing $\delta_0$ by a few percent we can change the amplitude of the power spectrum by more than 200 orders of magnitude. Recall that the PBH abundance is exponentially sensitive to $\PRfrozen$.
The results are the same in the Palatini case, except for the maximum value of $\xi$, which is 22 000 instead of 90, and the range of $\delta_0$, which is $0.03\dots0.35$ instead of $1.48\dots1.55$.

The allowed parameter space is bounded as follows: at smaller $\delta_0$ values and bigger $\xi$ values, $N$ corresponding to the correct $A_s$ is always too large; at bigger $\delta_0$ values, $N$ is always too small; and at lower $\xi$ values, $n_s$ is always too small (the values shown in figure \ref{fig:scans_big_d0_high_eta0} are cut at $n_s=0.9$). The maximum amplitude of $\PRfrozen$ rapidly varies from $<10^{-4}$ to $\gg1$.

\begin{figure}
\begin{center}
\includegraphics{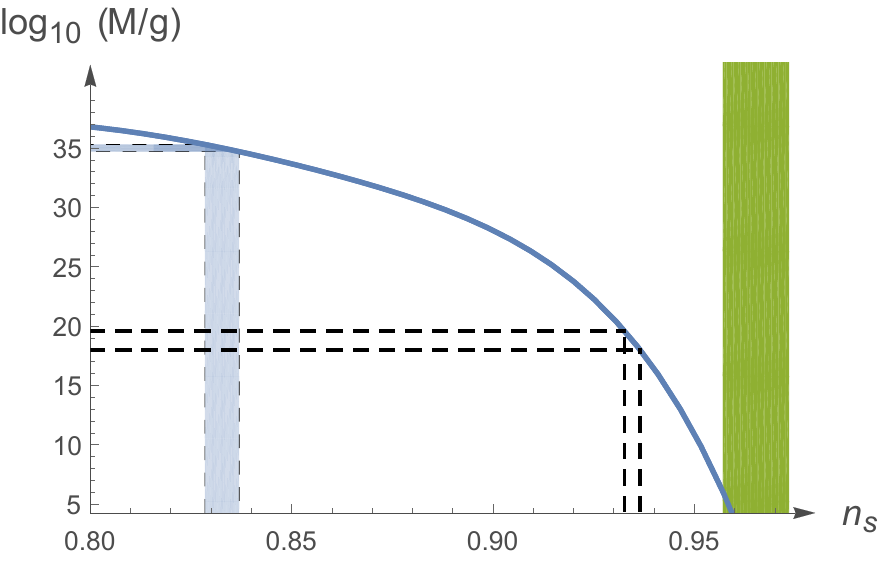}
\end{center}
\caption{PBH mass $M$ plotted against the CMB spectral index $n_s$ with the data of figure \ref{fig:scans_big_d0_high_eta0}. Dashed lines show the observationally allowed PBH mass windows ($10^{18}$ g, $4\times10^{19}$ g, and $10^{34}\dots10^{35} \, \mr{g} \approx 25\dots100 \, \mr{M}_\odot$), and the corresponding values for $n_s$. The green shaded area corresponds to the Planck 95\% confidence limits \eqref{eq:ns_obs} on $n_s$. This figure is for the metric case; the behaviour is similar in the Palatini case.}
\label{fig:M_in_ns}
\end{figure}

We also have to get the perturbations right in the range probed by CMB observations. As we see from figure \ref{fig:scans_big_d0_high_eta0}, the PBH mass and the spectral index are tightly correlated. We show this more clearly in figure \ref{fig:M_in_ns}. As the PBH mass increases, the spectral index $n_s$ at the pivot scale decreases, moving away from the observed value $0.9653\pm0.0041$. We have indicated the observational mass windows where PBHs can form a significant fraction of dark matter \cite{Carr:2017jsz}. The LIGO scale mass window $M\approx10^{34}\dots10^{35} \, \mr{g} \approx 25\dots100 \, \mr{M}_\odot$ of solar mass black holes is excluded, as $n_s<0.85$. The lower mass windows $M\approx10^{18}$ g and $M\approx4\times10^{19}$ g give $n_s\approx0.93\dots0.94$, over 6$\sigma$ from the observed value. While different datasets give slightly different values for the spectral index, all are in strong tension with such a low $n_s$ \cite{Akrami:2018odb}.

The scenario with Planck mass relics as dark matter requires, according to \eqref{eq:PBHmass_upper_limit_to_not_dominate}, $M<10^6$ g. Figure \ref{fig:M_in_ns} shows that the spectral index agrees with the observations precisely for this mass range. Since the peak in the power spectrum is formed at scales that exit the Hubble radius near the end of inflation, there is no SR plateau after the USR region, and the peak in the power spectrum is very sharp, see figure \eqref{fig:PR_in_lnk_peak}. There is no room for a long tail in the power spectrum after the feature, since inflation ends soon after the Hubble exit of the peak scale, and after inflation, the power spectrum is suppressed. By tuning the parameters, any amount of relics can be produced without violating the observational limits, so they can constitute all of dark matter. This corresponds to $\xi=70\dots90$ in the metric formulation and $\xi=(5\dots 22)\times10^3$ in the Palatini formulation, much smaller than in tree-level plateau inflation, where $\xi \sim 10^4$ (metric) and $\xi \sim 10^{9}$ (Palatini) (nelecting the running of $\lambda$ from the EW scale to the inflationary scale). This may alleviate possible problems with unitarity \cite{Barbon:2009ya, Burgess:2009ea, Hertzberg:2010dc, Bauer:2010jg, Bezrukov:2010jz, Bezrukov:2011sz, Calmet:2013hia, Weenink:2010rr, Prokopec:2012ug, Xianyu:2013rya, Prokopec:2014iya, Ren:2014sya, Escriva:2016cwl} and sensitivity to initial conditions \cite{Salvio:2015kka, Salvio:2017oyf}.

With Planck mass relics, the spectrum in the CMB region is the same as in tree-level plateau inflation, with $n_s=0.96$, $r=5\times10^{-3}$ (metric) and $r=4\times 10^{-8} \dots 2 \times 10^{-7}$ (Palatini).\footnote{In the tree-level Palatini case, $r=5\times10^{-3}/(6\xi)$, so $r$ can take various values depending on the value of the $\xi$, which is determined by the value of $\lambda$ at the pivot scale. Our value for $r$ is the same as in the tree-level case for the corresponding value of $\xi$.} The running and running of the running of the spectral index are also identical, as the inflationary plateau has the same shape as in the tree-level case. The predictions agree with the tree-level case for dozens of e-folds around the pivot scale. This is different from the case when the critical point is in the inflationary region, when $r$ can be much larger \cite{Allison:2013uaa, Bezrukov:2014bra, Hamada:2014iga, Bezrukov:2014ipa, Rubio:2015zia, Fumagalli:2016lls, Enckell:2016xse, Bezrukov:2017dyv, Rasanen:2017ivk, Salvio:2017oyf, Masina:2018}. However, quantum corrections do have a large impact on the Higgs quartic coupling $\lambda$. At the EW scale, $\lambda=0.13$, but on the inflationary plateau $\lambda$ is actually negative. However, the effective quartic coupling defined as $U/(\frac{1}{4} F^4)$ (where $F$ is defined in \eqref{eq:F_metric} and \eqref{eq:F_palatini}) is positive, as the quantum corrections are significant. At the pivot scale, the effective coupling is $\lambda=(2\dots5)\times10^{-6}$ (metric) or $\lambda=(0.5\dots3)\times10^{-6}$ (Palatini), corresponding to the amplitude of perturbations \eqref{eq:pert_norm} being proportional to $\lambda/\xi^2$ (metric) or $\lambda/\xi$ (Palatini).

\subsubsection{Jumps in $\lambda$ and $y_t$} \label{sec:jumps}

In the parameter region of figure \ref{fig:scans_big_d0_high_eta0} in the metric case, the jumps in $\lambda$ and $y_t$ are $\Delta\lambda\approx0.008$, $\Delta y_t\approx-0.02$ for the mean EW mass values $m_H$ and $m_t$ given in \eqref{eq:SM_bestfit_vals}\footnote{The SM value of $\lambda$ is actually negative at the threshold \eqref{eq:jumpscale} for the mean values of the masses, but this is not necessarily a problem since finite temperature corrections can make the potential positive \cite{Bezrukov:2014ipa, Rubio:2015zia, Enckell:2016xse}, and the results are anyway not sensitive to changing the EW scale parameters, as explained in the text.}. Since the jumps decouple EW and inflationary scale physics, the EW scale masses can be changed with practically no effect on the inflationary observables, as the difference is absorbed by the jumps. However, to make the jumps zero in the region of parameters where Planck-scale relics constitute all of the dark matter, we need $m_H=119\dots121$ GeV and $m_t=168\dots169$ GeV in the metric case, depending on the precise parameter values. Compared to the experimental values \eqref{eq:SM_bestfit_vals}, $m_H$ is more than 16$\sigma$ below the mean, and $m_t$ is 3 to 7$\sigma$ below the mean, depending on the uncertainty in connecting the experimental Monte Carlo mass to the theoretical pole mass \cite{Khachatryan:2015hba, Butenschoen:2016lpz}. Thus the jumps are required to form a critical or near-critical point, at least in the interesting parameter region considered here. This is also true when the critical point is in the CMB region \cite{Bezrukov:2014bra, Bezrukov:2014ipa,  Rubio:2015zia, Enckell:2016xse, Bezrukov:2017dyv}, in contrast to hilltop inflation, where successful hilltop inflation can happen in the CMB region with zero jumps \cite{Enckell:2018kkc}.

In the Palatini case, to make the jumps zero with Planck relics constituting all dark matter, we would need $m_H=114\dots117$ GeV and $m_t=165\dots167$ GeV, both far from the mean values \eqref{eq:SM_bestfit_vals}.

\section{Discussion} \label{sec:discussion}

\subsection{Comparison to previous work}

PBH production in Higgs inflation was first studied in \cite{Ezquiaga:2017fvi} in a simple approximation where the running of the quartic Higgs coupling and of the non-minimal coupling are fit phenomenologically. The authors concluded that Higgs inflation could produce all of dark matter as PBHs in the mass range $10^{17}\dots10^{21}$ g. Their analysis used the SR approximation, and involved significant running of $\xi$, which was criticized in \cite{Bezrukov:2017dyv, Germani:2017bcs}. We have shown that the SR approximation can fail near the critical point (although it may give the correct order of magnitude for the power spectrum if there is no local minimum in the potential, see figure \ref{fig:PR_in_N}), and have checked that the running of $\xi$ is negligible. When this is properly taken into account, we find that generating a significant amount of PBHs in this mass window leads to $n_s < 0.94$ at the CMB pivot scale, incompatible with the CMB results, as shown in figure \ref{fig:M_in_ns}. Our conclusion agrees with those of \cite{Motohashi:2017kbs, Germani:2017bcs, Bezrukov:2017dyv}.

Our results are similar to those of \cite{Ballesteros:2017fsr}. The authors studied a critical point model with non-minimal coupling to gravity and quantum corrections similar to Higgs inflation. The main difference is that we have considered a set of realistic quantum corrections for the Higgs field \eqref{eq:one_loop_corr} with a connection to EW scale collider physics and a carefully chosen renormalization scale \eqref{eq:renormscale_chiral} instead of generic correction terms, and we have scanned over the whole space of allowed critical point potentials. The studies share a strong correlation between PBH mass $M$ and the pivot scale spectral index $n_s$, with $n_s$ decreasing with increasing $M$. Using a shallow minimum instead of an exact critical point during USR to enhance the power spectrum was also suggested in \cite{Kannike:2017bxn, Ballesteros:2017fsr, Bezrukov:2017dyv}. We found that this is a required feature for Higgs inflation to produce a significant amount of PBHs.

In paper \cite{Masina:2018}, it was concluded that critical Higgs inflation is not a viable scenario. However, the unknown physics at scales $1/\xi < h < 1/\sqrt{\xi}$, which we encode in the jumps of the couplings $\lambda$ and $y_t$, was neglected. In agreement with \cite{Masina:2018}, we also found in section \ref{sec:jumps} that near-critical point inflation is not viable with zero jumps, as has also been noted earlier in the case when the near-critical point is in the CMB region \cite{Bezrukov:2014bra, Bezrukov:2014ipa,  Rubio:2015zia, Enckell:2016xse, Bezrukov:2017dyv}.

\subsection{Quantum diffusion} \label{sec:diffusion}

In the above analysis, we have neglected the effect of quantum diffusion, where quantum fluctuations are large enough to affect the evolution of the local effective background field.
Naively, the importance of diffusion can be estimated by comparing the change in the inflaton field $\chi$ in one Hubble time induced by the classical rate of change $\dot{\chi}_\mr{cl}$ and the growth rate of its quantum fluctuations, $\expval{\delta \chi^2}\approx H^3 t/(4\pi^2)$ \cite{Vilenkin:1983xq}:
\begin{equation} \label{eq:diffusion_ratio}
	\frac{\sqrt{\frac{d\expval{\delta\chi^2}}{dt}/H}}{\dot{\chi}_\mr{cl}/H} \approx \sqrt{\frac{H^4}{4\pi^2\dot{\chi}_\mr{cl}^2}} \ ,
\end{equation}
which is equal to the SR approximation \eqref{eq:PR_in_de_sitter} for the power spectrum. We noted earlier that the true power spectrum is always larger than this in the USR phase, and it cannot exceed unity so as not to overproduce PBHs, so the ratio \eqref{eq:diffusion_ratio} is always small in the parameter region we are interested in. This suggests that quantum diffusion is not important in the scenario we consider, and that we also do not need to worry about the ambiguities of eternal inflation.

However, see \cite{Biagetti:2018pjj} for criticism of this naive criterion for stochastic effects to be important, and \cite{Pattison:2017mbe,Ezquiaga:2018gbw,Cruces:2018cvq, Pattison:2018, Pinol:2018} for further discussion of quantum diffusion at a critical point and beyond SR and its effects on PBH formation. In particular, it was argued in \cite{Ezquiaga:2018gbw} that quantum diffusion can greatly enhance the power spectrum at the feature scale; however, this was questioned in \cite{Cruces:2018cvq}. Details of the effect of quantum diffusion on PBH formation are still under debate. Because in our case tuning the parameters (in particular, $\delta_0$) can change $\PRfrozen$ by many orders of magnitude while leaving the CMB observables unaffected, it seems that quantum diffusion would not affect our conclusions regarding Planck scale relics in critical point Higgs inflation.

\subsection{Non-Gaussianities} \label{sec:nongaussianities}

PBHs are formed from extreme perturbations in the tail of the probability distribution \eqref{eq:gaussiandistribution}, so their production can be sensitive to non-Gaussianities \cite{Franciolini:2018vbk, Ezquiaga:2018gbw}. It is therefore not clear how reliable our results for PBH abundance based on a Gaussian treatment, such as \eqref{eq:omegaPBH} are. Also, if we take PBHs to form on peaks of the density distribution, their distribution is not Gaussian, and this may change the PBH abundance \cite{Green:2004wb, Young:2014ana, Yoo:2018, Germani:2018jgr}. Simply calculating the width of the distribution more accurately with a window function for a Gaussian distribution can also have a significant effect on the abundance \cite{Ando:2018}.

However, as explained above, a small change in the input parameters can change the abundance of Planck-scale relic PBHs by orders of magnitude, so some modifications to the shape of the tail of the distribution can be countered by simply changing our input parameters. We therefore expect our result for Planck-scale relics to be robust against a range of such corrections. However, significant enhancement of the power spectrum could possibly change our conclusion that it is not possible to form a sufficient amount of PBHs with only a critical point and a shallow minimum is needed. Since we found that the maximum value of $\PRfrozen$ from a critical point compatible with CMB observations is $10^{-6}$, an enhancement of at least factor 100 would be needed to produce a significant amount of PBHs.

\subsection{Critical point with USR on CMB scales}

For PBHs to constitute the dark matter, the feature that produces them must be below the CMB scale. However, PBH production aside, we can ask what kind of observational signatures a critical point with USR in the CMB region could produce in the power spectrum.

Since in this case inflation has to last for 50 e-folds after the Hubble exit of the feature scale, there must be a long SR period after the USR period. The power spectrum then resembles that of figure \ref{fig:PR_in_N}: two SR plateaus with a step-like USR transition in between. It is not possible to obtain a sharp peak as in figure \ref{fig:PR_in_lnk_peak}. Features of this kind were considered in \cite{Starobinsky:1992ts}. In principle, if the height of the step could be adjusted freely while keeping the step sharp, it might provide an interesting feature in the observed CMB spectrum.

Sharpness is key: a sharp step can't be produced in SR, and may be hard to detect directly with observational analyses that assume approximate scale-independence. However, the height and the sharpness of the step are correlated. If the SR conditions are violated strongly near the critical point, then the transition to USR is quick, and USR lasts long before the field evolution returns to SR. This produces a high step in $\PRfrozen$. If the SR conditions are violated less, then USR does not last as long and the step is lower, but $\PRfrozen$ grows more already in the SR phase as $\epsilon_H \to 0$ (see \eqref{eq:PR_in_SR}), and the step is not so sharp. To produce a sharp and low step, the potential must have a very special form. In Higgs inflation, the height of a sharp step is always many orders of magnitude, too large to agree with CMB observations.

\section{Conclusions} \label{sec:conclusions}

We have studied the possibility that primordial black holes (PBH) formed in critical point Higgs inflation make up part or all of the dark matter. Near a critical point, where the first and second derivatives of the potential vanish, the slow-roll approximation may break down, and the inflaton may enter a period of ultra-slow-roll. In ultra-slow-roll the comoving curvature perturbation can be significantly enhanced, even on super-Hubble scales.

We consider the Standard Model Higgs with a non-minimal coupling $\xi$ to gravity. We take into account three-loop quantum corrections to the potential in the small field limit, and consider the chiral Standard Model with one-loop corrections in the large field limit. We match them at an intermediate field value with arbitrary jumps in the Higgs quartic coupling and the top Yukawa coupling, demanding that the quantum corrections generate a near-critical point. In addition to the jumps, we have $\xi$ and the location of the near-critical point as free parameters.

We scan over all potentials with a near-critical point by adjusting these four parameters, calculate the maximum value of the asymptotic power spectrum $\PRfrozen$ by numerically solving the mode equations and check if the right amount of PBHs are formed to constitute dark matter, while satisfying the observational constraints for PBHs and the CMB power spectrum. We consider both the metric and the Palatini formulation of general relativity.

We find that if the potential is monotonic, a near-critical point cannot produce noticeable amounts of PBHs. We then consider potentials with a shallow local minimum instead. As the field rolls up from the minimum, it slows down so much that the SR approximation fails by orders of magnitude. We find that $\PRfrozen$ is enhanced for scales that exit the Hubble radius as the field rolls up. This mechanism was first introduced in \cite{Kannike:2017bxn, Ballesteros:2017fsr, Bezrukov:2017dyv}.

The resulting PBH mass $M$ and the spectral index $n_s$ on CMB scales are highly correlated: as $M$ increases, $n_s$ decreases and moves away from the observed value of $0.965$. For $n_s$ to be compatible with the CMB observations, $M$ has to be so small that the PBHs would evaporate before BBN. This is contradiction with a previous study on PBH production in Higgs inflation where the authors considered only the SR approximation and a phenomenological form for the Higgs potential with a large running of the non-minimal coupling $\xi$ \cite{Ezquiaga:2017fvi}.

However, if PBHs do not evaporate completely but leave behind Planck mass relics, early evaporation is not a problem: apart from the relic density, there are no observational constraints on PBHs that evaporate before the EW crossover and do not spoil baryogenesis. Such PBHs have initial masses $M<10^6$ g. It is intriguing that they correspond to scales that exit $50$ e-folds after the exit of the CMB pivot scale, so the feature is at the place in the potential where inflation ends. The abundance of such relics can be adjusted freely by tuning the parameters of the model. In particular, it can be matched to the observed amount of dark matter. These conclusions hold both for the metric and the Palatini case.

In this mass range the spectral index is also exactly right for observations. The CMB observables agree with the tree-level predictions of Higgs inflation \cite{Bezrukov:2007ep, Bauer:2008}: $n_s=0.96$ and $r=5\times10^{-3}$ (metric), $r=4\times 10^{-8} \dots 2 \times 10^{-7}$ (Palatini). However, the non-minimal coupling is much smaller, $\xi=70\dots90$ (metric) or $\xi=(5\dots22)\times10^3$ (Palatini) than in the tree-level case where $\xi=10^4$ (metric) or $\xi=10^9$ (Palatini). (The effective Higgs quartic coupling is also correspondingly smaller.) Therefore, perturbative unitarity is violated at a higher scale than in the tree-level case, which may alleviate possible problems with unitarity \cite{Barbon:2009ya, Burgess:2009ea, Hertzberg:2010dc, Bauer:2010jg, Bezrukov:2010jz, Bezrukov:2011sz, Calmet:2013hia, Weenink:2010rr, Prokopec:2012ug, Xianyu:2013rya, Prokopec:2014iya, Ren:2014sya, Escriva:2016cwl} and sensitivity to initial conditions \cite{Salvio:2015kka, Salvio:2017oyf}.

One caveat is that for the large field values we have used the chiral SM, which strictly speaking only applies for $\delta\ll1$. This condition holds at the CMB scales, but on the scale where the PBHs are produced we have $\delta\approx1.5$ in the metric case, so our approximation for the potential may not be valid. In the Palatini case we have $\delta=0.03\dots0.35$ instead, so the approximation may be better under control. However, for finite $\delta$ the differences in the renormalization group running between the metric and the Palatini case should also be taken into account.

We also did not consider the effects of quantum diffusion and non-Gaussianities. Looking at PBH formation in more detail with the peaks formalism and a proper window function could also change the abundance of PBHs. However, small changes in our input parameters can change the power spectrum $\PRfrozen$ by orders of magnitude without effect on the CMB observables. Therefore, while the conclusion that a critical point cannot produce enough PBHs could be changed by such corrections, we expect our result that Planck scale PBHs produced in Higgs inflation can be the dark matter to be robust against a range of such corrections. This would mean that two phenomena ---inflation and dark matter--- that are usually taken as evidence for physics beyond the Standard Model could be explained with no new particle physics.

\acknowledgments

ET is supported by the Vilho, Yrj{\"o} and Kalle V{\"a}is{\"a}l{\"a} Foundation of the Finnish Academy of Science and Letters.

\appendix

\section{Upper mass limit for Planck scale relic PBHs} \label{app:uppermasslimit}

As discussed in section \ref{sec:relics}, when PBHs evaporate they may leave behind Planck mass relics that could constitute dark matter. In this appendix, we find the maximum initial mass $M$ for such PBHs, assuming they all have the same mass. The PBHs form in  the early radiation-dominated era. They form with negligible initial momentum, so they behave as cold dark matter. Hence their energy density decreases slower than the energy density of radiation, so their fractional contribution $\Omega_\mr{PBH}$ to the total energy density grows. However, evaporation slowly turns their energy back into radiation, and as the evaporation nears completion, $\Omega_\mr{PBH}$ plummets and only a small fraction is left as the relic energy density $\Omega_\mr{rel}$, which then again grows and eventually overtakes the radiation.

There are stringent limits for black hole evaporation and energy density fraction, and these set an upper limit for the initial mass of PBHs if the relics are to constitute dark matter. In particular, evaporation should be finished before  BBN so as not to spoil its predictions \cite{Carr:2009jm, Carr:2017jsz}. Let us calculate the limit this gives on the mass.

The lifetime of a PBH with initial mass $M$ is \cite{MacGibbon:1991tj}
\begin{equation} \label{eq:BH_lifetime}
	t_\mr{ev} = \frac{1.88}{f_L(M)} \qty(\frac{M}{\Mpl})^3 \Mpl^{-1}
\end{equation}
where $f_L(M)$ depends on the number, charge and spin of the effectively massless degrees of freedom into which the PBH evaporates during its lifetime. For the small masses we consider, $f_L(M)\approx13.9$.

During radiation domination, the Friedmann equation can be written in terms of the temperature $T$ as
\begin{equation} \label{eq:friedmann_temp}
	3H^2 = \frac{\pi^2}{30} g_*(T)  \frac{T^4}{\Mpl^2} \ ,
\end{equation}
where $g_*(T)$ is the effective number of degrees of freedom in the SM. Using $H=1/(2t)$, the equations \eqref{eq:BH_lifetime} and \eqref{eq:friedmann_temp} give the temperature at the end of evaporation:
\begin{equation} \label{eq:ev_temp}
	T_\mr{ev} = \frac{0.896 f_L^{1/2}}{g_*^{1/4}} \qty(\frac{\Mpl}{M})^{3/2} \Mpl \ .
\end{equation}
Using $T_\mr{ev}>4.7$ MeV as the limit on the temperature at which extra radiation can be injected without affecting BBN \cite{deSalas:2015}, for which $g_*(T)\geq10.75$, we get
\begin{equation} \label{eq:M_BBN_limit}
	M < 4.1 \times 10^8 \ \mr{g} \quad \text{for PBHs to evaporate before BBN.}
\end{equation}

We will next derive an even stronger limit for $M$ based on arguments regarding PBHs dominating the energy density of the universe at some point in its history. We proceed in three steps. First, we derive upper and lower limits for $M$ in the scenario where PBHs are assumed to dominate at some point. Second, we observe that these limits contradict each other, so PBH domination can't occur. Third, based on this observation, we derive an upper limit on $M$. 

We start with the constraint that PBHs must not dominate the energy density after the EW crossover at $T_\mr{EW} = 160$ GeV \cite{DOnofrio:2015}, since this would wipe out any baryon asymmetry, which presumably cannot be generated after the EW crossover \cite{Carr:1994ar}\footnote{There are some mechanisms of baryogenesis which circumvent this requirement. For example, in some extensions of the SM baryon asymmetry could be generated by black holes evaporating into heavy bosons whose decay generates the asymmetry \cite{Barrow:1980, Barrow:1990qt, Barrow:1990he}.}. Thus, in the scenario where PBHs dominate, they must evaporate before the crossover, that is, $T_\mr{ev} > T_\mr{EW}$ (for which $g_*(T)\geq96.25$), which gives
\begin{equation} \label{eq:M_electroweak_limit}
	M < 2.7 \times 10^5 \ \mr{g} \quad \text{for PBHs to evaporate before $T_\mr{EW}$.}
\end{equation}
This was calculated in a radiation-dominated background; if there is a period of PBH domination, the limiting mass in \eqref{eq:M_electroweak_limit} will be lower, since during matter domination, temperature decreases faster and only PBHs with even smaller masses have time to evaporate before $T_\mr{EW}$ is reached.

By demanding that PBHs are not overproduced, we can also derive an upper limit for $M$ in the scenario where PBHs dominate at some point before the EW crossover. The relic energy density fraction at matter-radiation equality can be written as
\begin{equation} \label{eq:omega_relic}
	\Omega_\mr{rel\,eq} = \Omega_1 \frac{M_\mr{rel}}{M_1} \frac{ T_1 g_{*S} (T_1)^{1/3} }{ T_\mr{eq} g_{*S} (T_\mr{eq})^{1/3} } \ ,
\end{equation}
where $\Omega_1$ is the energy density fraction of PBHs at some moment $t_1$ after matter domination but before the end of evaporation --- let us take $\Omega_1=1/2$. The first ratio $M_\mr{rel}/{M_1}$ gives the decrease from the mass $M_1$ at $t_1$ evaporating down to the relic mass $M_\mr{rel}$. The second ratio gives the increase of $\Omega_\mr{rel}$ due to PBH energy density rising like $a$ relative to radiation. Demanding that PBHs dominate before the EW transition, $T_1 > T_\mr{EW}$, that $\Omega_\mr{rel\,eq} < 1/2$ so that dark matter is not overproduced and setting $M_\mr{rel} = \Mpl$ and using $g_{*S}(T_\mr{EW})=96.25$, $T_\mr{eq}=0.79$ eV, $g_{*S}(T_\mr{eq})=3.909$,
\eqref{eq:omega_relic} gives for the initial PBH mass:
\begin{equation} \label{eq:M_lower_limit_when_dominating}
	M > M_1 > 2.6\times10^6 \ \mr{g} \quad \text{for PBHs to dominate without overproducing relics.}
\end{equation}
This is in contradiction with the limit \eqref{eq:M_electroweak_limit}. We conclude that if PBHs leave behind Planck mass relics, then PBHs can't dominate at any time or they either spoil baryogenesis or produce too much dark matter.

Finally, the requirement that PBHs are always subdominant can be translated into a limit on their initial mass, if we also demand that they contribute considerably to dark matter. From \eqref{eq:BH_lifetime}, mass of a black hole as a function of time is
\begin{equation} \label{eq:M_in_time}
	M(t) = M\qty(1-\frac{t}{t_\mr{ev}})^{1/3} \ ,
\end{equation}
so before the end of evaporation, in a radiation dominated universe, we have
\begin{equation} \label{eq:omega_rel_in_time}
	\Omega_\mr{rel}(t) \propto a(t)M(t) \propto \qty(\frac{t}{t_\mr{ev}})^{1/2}\qty(1-\frac{t}{t_\mr{ev}})^{1/3} \ ,
\end{equation}
which has a maximum at $t_\mr{m} = \frac{3}{5}t_\mr{ev}$, with $M(t_\mr{m}) = (2/5)^{1/3}M$. Using \eqref{eq:omega_relic} with $t_1=t_\mr{m}$ gives, using $T g_{*S}(T)^{1/3}\propto a^{-1}\propto t^{-1/2}$, applying \eqref{eq:BH_lifetime} and taking $g_{*}(T_\mr{ev})=g_{*S}(T_\mr{ev})=106.75$ (using 96.25 would give the same result, as the dependence on $g_{*S}(T_\mr{ev})$ is weak, $M\propto g_{*S}(T_\mr{ev})^{1/30}$),
\begin{equation} \label{eq:M_solved}
	M = 8.5 \times 10^5 \qty(\frac{\Omega(t_\mr{m})}{\Omega_\mr{rel\,eq}})^{2/5} \mr{g} \ .
\end{equation}
For $\Omega(t_\mr{m}) < 1/2$ (no PBH domination) and $\Omega_\mr{rel\,eq}>0.1$ (considerable dark matter fraction in relics) this gives
\begin{equation} \label{eq:PBHmass_upper_limit_to_not_dominate}
	M < 1.6 \times 10^6 \ \mr{g} \qquad
	\begin{array}{l}
	\text{for relics to significantly contribute to DM} \\
	\text{without PBH domination.}
	\end{array}
\end{equation}
This result is not unique to Higgs inflation but applies to all cases where PBHs form at a single mass scale and baryogenesis happens at $T>T_\mr{EW}$. It comes with the caveat that we have assumed $M_\mr{rel} = \Mpl$. If the relic mass is a few orders of magnitude smaller, the limits \eqref{eq:M_electroweak_limit} and \eqref{eq:M_lower_limit_when_dominating} overlap and PBH domination could be possible, but a more detailed calculation in a non-radiation dominated background is needed to determine this. PBH domination could have specific observational signatures such as gravitational waves \cite{Dolgov:2011}.

\section{Growth of $\PR$ after Hubble exit} \label{app:PRgrowth}

According to the usual lore, the comoving curvature perturbation $\R_k$ and thus the power spectrum $\PR(k,t)$ freeze to a constant value after Hubble exit ($aH \gg k$) if perturbations are adiabatic. However, in USR inflation, this does not happen \cite{Kinney:2005vj}. In this appendix, we explain why.

The general proof of the freezing of $\R$ goes as follows \cite{Lyth:2004gb}. First, we write down the local energy continuity equation in the gradient expansion
\begin{equation} \label{eq:local_energy_conservation}
	\dot{\rho}(\vec{x},t) = -3\qty[\frac{\dot{a}(t)}{a(t)} + \dot{\psi}(\vec{x},t)] \qty[\rho(\vec{x},t) + p(\vec{x},t)] + O(\epsilon^2) \ ,
\end{equation}
where $\psi$ is the curvature perturbation, and $\epsilon \equiv k/(aH)$. Equation \eqref{eq:local_energy_conservation} without the $O(\epsilon^2)$ terms supposedly applies on the largest, super-Hubble scales with $\epsilon \ll 1$, where the universe is `locally homogeneous and isotropic' with a local Hubble parameter $\widetilde{H}\equiv\dot{a}/a + \psi$.

Second, we go to the uniform-density gauge where $\rho(\vec{x},t)=\rho(t)$ is spatially constant. If the equation of state is barotropic, that is, pressure $p$ is a function of energy density $\rho$ only\footnote{This assumption holds for a single scalar field, for which $\rho = \frac{1}{2}\dot{\chi}^2 + V(\chi)$, $p = \frac{1}{2}\dot{\chi}^2 - V(\chi)$ at the homogeneous limit, if initial conditions are fixed. This fixing is done by SR inflation. Even if SR approximation is later violated, the barotropic condition still holds, since $p$ can be deduced from $\rho$ that decreases monotonically in way determined by the initial SR period. For the proof of the conservation of $\R$ in the scalar field case, see section 2.2 of \cite{Gordon:2001}.}, then $p$ is also spatially constant, and so is $\dot{\psi}$, up to corrections $O(\epsilon^2)$. But since $\psi$ is a perturbation whose mean value should be zero, we have $\dot{\psi}(t)=0$ and $\psi(\vec{x},t)$ is constant in time. (The perturbation $\psi$ is usually denoted by $\zeta$ in the uniform-density gauge.)

Third, we notice that on super-Hubble scales, the comoving and uniform density gauges coincide, so the curvature perturbation $\R$ in the comoving gauge is also constant.

A crucial assumption in the above analysis is that terms of order $\epsilon^2$ can be omitted as a small quantity. However, in USR, we have for the inflaton field $\chi$
\begin{equation} \label{eq:gradient_expansion_breaking}
	\ddot{\chi}+ 3H\dot{\chi} = 0 \quad \Rightarrow \quad \dot{\chi} \propto e^{-3H} \propto a^{-3} \quad \Rightarrow \quad p+\rho = \dot{\chi}^2 \propto a^{-6} \ ,
\end{equation}
that is, if $\dot{\psi}=0$, $\widetilde{H}(p + \rho)$ decreases exponentially and much faster than $\epsilon^2 \propto a^{-2}$. Thus, for any non-zero $k$, the $O(\epsilon^2)$ terms become comparable to the leading terms in a finite time, and the gradient expansion breaks down. The usual result that $\R$ freezes on super-Hubble scales is therefore not applicable.

Indeed, if we write down the relation between $\zeta_k$ and $\R_k$ or the time evolution equation of $\R_k$ at lowest order in perturbation theory \cite{Mukhanov:1990me}, we encounter ratios such as $k^2/[a^2H^2(1 + p/\rho)]$, which are usually assumed to be small on super-Hubble scales, but which in USR grow exponentially in time. This is not the case in SR: there $\dot{\chi}$ is small but approximately constant, as required by the SR condition $|\eta_H| < 1$. For $H\approx$ constant, $\eta_H\approx$ constant, we have $\dot{\chi}^2 \propto a^{-2\eta_H}$, which decreases slower than $\epsilon^2$. For further discussion on the super-Hubble behaviour in such constant-roll inflation, see \cite{Motohashi:2014ppa, Yi:2017mxs}. (Note that USR is a singular limit of constant-roll \cite{Pattison:2018}.)

\bibliographystyle{JHEP}
\bibliography{inflp}

\end{document}